\documentclass[journal]{IEEEtran}
\IEEEoverridecommandlockouts
\usepackage[utf8]{inputenc}
\usepackage{amsmath,amsfonts}
\usepackage{algorithmic}
\usepackage{algorithm}
\usepackage{array}
\usepackage{hyperref}
\ifCLASSOPTIONcompsoc   
\usepackage[caption=false,font=normalsize,labelfon
t=sf,textfont=sf]{subfig}
\else
\usepackage[caption=false,font=footnotesize]{subfi
g}
\fi
\usepackage{textcomp}
\usepackage{stfloats}
\usepackage{url}
\usepackage{verbatim}
\usepackage{graphicx}
\usepackage{cite}
\usepackage{xspace}
\usepackage{xcolor}
\usepackage{xargs}
\usepackage[colorinlistoftodos
	, prependcaption
	, textsize=footnotesize
	, disable
	, textwidth=2.85cm
	]{todonotes}
\newcommandx{\maybe}[2][1=]{\todo[color=orange!30!white,#1]{#2}}
\newcommandx{\note}[2][1=]{\todo[color=blue!50!white,#1]{#2}}
\newcommandx{\review}[2][1=]{\todo[color=lime!50!white,#1]{#2}}
\newcommandx{\wip}[2][1=]{\todo[color=purple!50!white,#1]{#2}}
\newcommand{\revised}[1]{%
{#1}}
\newenvironment{revisedenv}
{\color{black}}
{\ignorespacesafterend}
\newcommandx{\forCameraReady}[2][1=]{\todo[color=green!30!white,#1]{#2}}
\usepackage{tikz}
\usetikzlibrary{positioning,shapes,arrows}
\usepackage{listings}
\usepackage{multirow}
\graphicspath{{figures/}}
\definecolor{myDarkGreen}{rgb}{0.0,0.5,0.0}

\lstdefinelanguage[riscv]{Assembler}%
{
morestring=[b]",
morekeywords = [1]{.alias,.align,.ascii,.asciiz,.byte,.data,.double,.end,.endb,%
  .endr,.ent,.err,.extern,.file,.float,.fmask,.frame,.globl,.half,.text,%
  .verstamp,.vreg,.word},%
morekeywords = [2]{lw,sw,lb,lbu, lh, lhu, sb, sh, la,li,lui, ori, or, sub, subi, add, addi, or, ori, xor, xori, sra, srai, srl, srli, slli, srl, sll, beq, bne,bge,blt, bltu, bgeu,j,jal,jalr, jr, j, or,addi,ecall,ret,load_patch, call, push, pop, jmp},%
morekeywords = [3]{x0,x1,x2,x3,x4,x5,x6,x7,x8,x9,x10,x11,x12,x13,x14,x15,x16,x17,x18,x19,x20,x21,ra,s0,s1,s2,a0,a1,x28,x29,x30,x31,sp,zero, a0,a1,a2,a3,a4,a5,a6,a7,sp,ra,t0,t1,t2,t3,t4,t5,t6,t7,s0,s1,s2,s3,s4,s5,s6,s7,s11},%
comment = [l]\#,%
keywordsprefix=.,%
sensitive=false,%
}[keywords,comments,strings]

\lstdefinestyle{customriscv}{
  language = [riscv]Assembler,
  basicstyle = \footnotesize\ttfamily,
  stringstyle = \footnotesize\ttfamily,
  frame = single,
  rulesep = 5pt,
  numbers = none,
  framexleftmargin = \parindent,
  showspaces = false,
  showstringspaces = false,
  showtabs = false,
  rulecolor = \color{black},
  rulesepcolor = \color{black},
  tabsize=3,
  breaklines=false,
  keywordstyle = [1]\color{purple},%
  keywordstyle = [2]\color{blue},%
  keywordstyle = [3]\color{red},%
  stringstyle = \color{mauve},%
  escapeinside = {\%*}{*)},
  moredelim=[is][\bfseries\color{violet}]{;?}{?;},
  moredelim=[is][\bfseries\color{red}]{;!}{!;},
  moredelim=[is][\bfseries\color{green}]{;@}{@;},
  moredelim=[is][\bfseries]{[*}{*]},
  literate={à}{{\`a\ }}1 {é}{{\'e}}1 {è}{{\`e}}1 {ê}{{\^e}}1, %
  extendedchars=true
}

\newcommand{\code}[1]{\texttt{\small{}#1}}
\newcommand{\name}{MAFIA}

\begin{document}

\title{\name{}: Protecting the Microarchitecture of Embedded Systems Against Fault Injection Attacks\thanks{~}
}

\author{Thomas Chamelot, Damien Couroussé, Karine Heydemann
\thanks{This work was partially funded by the French National Research Agency (ANR) under grant agreement ANR-18-CE39-0003. 
Thomas Chamelot and Damien Couroussé are with the Univ. Grenoble Alpes, CEA, List, F-38000 Grenoble, France (e-mail: thomas.chamelot@cea.fr; damien.courousse@cea.fr). 
Karine Heydemann is with the Sorbonne Université, CNRS, LIP6, 75005 Paris, France and with
Thales DIS France
(e-mail: karine.heydemann@thalesgroup.com).
}
}

\maketitle

\forCameraReady[inline]{update des affiliations}

\begin{abstract}
Fault injection attacks represent an effective threat to embedded systems.
Recently, Laurent et al.\ have reported that fault injection attacks can leverage faults inside the microarchitecture.
However, state-of-the-art counter-measures, hardware-only or with hardware support, do not consider the integrity of microarchitecture control signals that are the target of these faults.

We present \name{}, a microarchitecture protection against fault injection attacks.
\name{} ensures integrity of pipeline control signals through a signature-based mechanism,
and ensures fine-grained control-flow integrity with a complete indirect branch support and code authenticity.
We analyse the security properties of two different implementations with different security/overhead trade-offs:
one with a CBC-MAC/Prince signature function, and another one with a CRC32.
We present our implementation of \name{} in a RISC-V processor, supported by a dedicated compiler toolchain based on LLVM/Clang.
{
  We report a hardware area overhead of 23.8\,\% and 6.5\,\% for the CBC-MAC/Prince and CRC32 respectively.
  The average code size and execution time overheads are 29.4\,\% and 18.4\,\% respectively for the CRC32 implementation and are 50\,\% and 39\,\% for the CBC-MAC/Prince.
}

\end{abstract}

\begin{IEEEkeywords}
fault injection attacks, code integrity, control-flow integrity, \revised{control-signal integrity}, code authenticity, control logic, counter-measures%
\end{IEEEkeywords}

\section{Introduction}

\IEEEPARstart{C}{\textbf{ontext.}}
{
Fault injection attacks are an important threat to the security of embedded systems~\cite{yuceFaultAttacksSecure2018}.
An attacker injects physical disturbances in a circuit, such as power or clock glitches, electromagnetic pulses, or laser beams, to induce a faulty behaviour.
This may result at the logical level in the alteration of several bits in different ways.
State-of-the-art attackers are able to control the alteration of one or few bit values~\cite{yuceFaultAttacksSecure2018, colombierMultiSpotLaserFault2022}.
}
The attacker aims at inducing computation errors or modifying values in the circuit under attack in order to leverage fault injection for many attack objectives such as the extraction of confidential data or privilege escalation. 

{
State-of-the-art counter-measures against fault injection attacks ensure three security properties: data integrity, code integrity, and control-flow integrity.
Data integrity ensures that data in storage, in transit or manipulated by the processor are not modified by any illegitimate means, e.g.\ by a fault inducing a bit-flip in a register.
Code integrity ensures that instructions of the program are not modified before their execution, for example a fault inducing a bit-flip in an instruction encoding.
Control-flow integrity ensures that the control-flow transfers, such as branches and calls, are correct with respect to a reference control-flow graph (CFG).
A full control-flow integrity also ensures the correct execution order of branchless instructions sequences, e.g.\ protects against a fault inducing an instruction skip.
All these properties are required to ensure the correct processing of a program.
}

Several works study code and control-flow integrity hardware mechanisms based on the computation of an integrity signature.
In~\cite{dangerCCFICacheTransparentFlexible2018}, a hardware monitor, external to the processor, computes a code integrity signature and uses additional metadata to validate the code and control-flow integrity in separate verification mechanisms.
In~\cite{wernerProtectingControlFlow2015}, a single signature mechanism ensures both code and control-flow integrity.
Finally, recent counter-measures for code and control-flow integrity are based on the authenticated decryption of program instructions~\cite{clercqSOFIASoftwareControl2016, wernerSpongeBasedControlFlowProtection2018, savryConfidaentControlFLow2020}.
{%
They also ensure code confidentiality and code authenticity in addition to control-flow integrity.
Note that code confidentiality prevents non-authorized entities from reading the program instructions thanks to encryption.
Code authenticity ensures that the binary program is emitted by an authorized entity, and if based on sound cryptographic mechanisms, also implies code integrity.
}

\textbf{Problem.}
Recently, Laurent et al.\ have reported that attacks can leverage faults inside the microarchitecture~\cite{laurentCrosslayerAnalysisSoftware2019}.
For example, a fault corrupting the write-back control signals after the decode stage will change the instruction behaviour.
State-of-the-art code and control-flow integrity counter-measures fail to catch such fault injection attacks because the fault does not modify the binary encoding of the instruction nor the control flow.
We argue that integrity of the control logic in the processor is required, together with data integrity, code integrity and control-flow integrity, to protect against fault injection attacks.
\begin{revisedenv}
We call \emph{control-signal integrity} the security property ensuring the integrity of the control logic in the processor.

\textbf{Goal \& Challenges.}
Our goal is to design a counter-measure against fault injection attacks simultaneously supporting control-flow integrity, code authenticity, and control-signal integrity.
Control-signal integrity 
protects the whole instruction path of the processor microarchitecture against fault injection attacks.
The first challenge is to implement a control-signal integrity mechanism, 
that is, to protect the whole control signals in the processor microarchitecture against fault injection attacks.
The second challenge is to combine control-signal integrity with a code and control-flow integrity approach that is robust against fault injection attacks.
Our last challenge is to implement the counter-measure in an embedded system 
with complete hardware and software support while maintaining a minimum overhead.

\textbf{Contributions.}
  This paper presents \name{}, the first counter-measure of our knowledge to ensure control-signal integrity against fault injection attacks, in combination with control-flow integrity and code authenticity.

\end{revisedenv}

\name{} is designed around the concept of \emph{pipeline state}, which is a selection of control signals representative of the current state of the processor.
An integrity signature is derived from the pipeline state, 
and any deviation from the expected signature values can be detected, highlighting a fault injection.
This approach ensures the integrity of all the control signals monitored {upstream from} the pipeline state.
	{Downstream from} the pipeline state, the monitored control signals are protected by a redundancy scheme, typical of counter-measures against fault injection attacks.
The combination of an integrity signature derived from the pipeline state with a redundancy-based protection ensures a full protection coverage of the control signals in the processor microarchitecture.

We detail the properties of the function signature required to ensure code integrity and control-flow integrity in our attacker model.
Code authenticity is also ensured when the function signature provides message authentication.

\name{} is extended with support for indirect control-flow transfers and interrupts,
which provides full support of software used in embedded systems.

\name{} is implemented as an extension of the CV32E40P RISC-V in-order processor, and is supported by a dedicated compiler toolchain.
We describe how \name{} is integrated to the processor architecture,
and we describe the modifications required for the compiler toolchain to fully support the counter-measure.

The signature function at the core of \name{} supports many possible implementations.
We evaluate two implementations with different security/overhead trade-offs:
one with a CBC-MAC integrating the Prince block cipher providing code authenticity,
and another one with a CRC32 error detection code providing code integrity only.
Notably, the integration of \name{} in the microarchitecture of the CV32E40P does not impact the design critical paths, allowing to maintain the target frequency of the reference ASIC implementation, at 400~MHz in the GF-22FDX FDSOI technology.
{%
  We report a hardware area overhead of 23.8\,\% and 6.5\,\% for CBC-MAC/Prince and CRC32 respectively.
  The average code size and execution time overheads are 29.4\,\% and 18.4\,\% respectively for CRC32 and are 50\,\% and 39\,\% for CBC-MAC/Prince.
}

{%
This paper is an extension of the work published in~\cite{chamelotSCIFIControlSignal2022}, in particular it presents support for indirect branches, branch prediction and interrupts.
It also gives more details regarding the hardware and software implementations, and it provides an analysis of \name{}'s security.%
}

\textbf{Outline.}
Section~\ref{sec:motivating example} illustrates why control-signal integrity is necessary.
Section~\ref{sec:background} introduces our threat model and then gives some background on code and control-flow integrity.
Section~\ref{sec:concept} details the design of \name{},
Section~\ref{sec:implementation} details our implementation.
Section~\ref{sec:security_analysis} provides a security analysis of \name{}, 
and Section~\mbox{\ref{sec:experimental_evaluation}} presents an evaluation of the resulting hardware and software overheads.
Finally,  Section~\ref{sec:related work} discusses related work  and Section~\ref{sec:conclusion} concludes.

\section{Motivating Example}
\label{sec:motivating example}
We illustrate the necessity of protecting the control signals in the microarchitecture with \revised{control-signal integrity},  and of combining this security property with code and control-flow integrity.
Listing~\ref{lst:example} is a small piece of RISC-V assembly code implementing a loop that exits when register \texttt{t0} equals \texttt{0}.

A single bit-flip applied on the binary encoding of the instructions, for example in program memory, could lead to the replacement of  instruction \texttt{bne} by an instruction \texttt{beq}, as illustrated in Listing~\ref{lst:example_fault}, leading to an inversion of the branch conditions.
Counter-measures ensuring code integrity would detect such a fault~\cite{aroraHardwareAssistedRunTimeMonitoring2006, wernerProtectingControlFlow2015, dangerCCFICacheTransparentFlexible2018, clercqSurveyHardwarebasedControl2017, wernerSpongeBasedControlFlowProtection2018, savryConfidaentControlFLow2020}.

If a fault with a similar branch inversion effect occurs in the microarchitecture during or after instruction decoding,
code integrity counter-measures fail to detect the fault because it does not modify the instruction encoding.
Moreover, regarding the control flow, the fault only appears as a branch inversion and does not alter the original program CFG. 
Therefore the fault can only be detected by control-flow integrity counter-measures tracking the integrity of branch conditions~\cite{SchillingSecuringConditionalBranches2018}.

Other faults in the microarchitecture can have harmful effects.
For example in Listing~\ref{lst:example}, 
a single bit-flit in the control signal of the forwarding mechanism can prevent forwarding of the \texttt{addi} instruction result in register \texttt{t0} to the \texttt{bne} instruction. 
If such a fault is injected during the last loop iteration, the \texttt{bne} instruction uses the previous value in \texttt{t0}, leading to an additional iteration instead of exiting the loop.
This is why ensuring \revised{control-signal integrity} in the microarchitecture is required to ensure the correct execution of a program.
Note 
that a fault applied before instruction decoding, i.e.\ into program memory or during instruction fetch, may be detected by code integrity but not by \revised{control-signal integrity}.
Therefore, it is necessary to ensure code, control-flow and \revised{control-signal integrity} and to cover the entire instruction path.

\begin{figure}
\begin{lstlisting}[captionpos=b,caption={Example of RISC-V instructions sequence implementing a loop.  Binary code on the left, assembly machine instructions on the right.}, label=lst:example, frame=single, language={[riscv]assembler}]
loop:
  ff f2 82 93   addi t0, t0, #-1
  fe 04 9e e3   bne t0, zero, loop
\end{lstlisting}

\begin{lstlisting}[captionpos=b,caption={Instructions sequence from Listing~\ref{lst:example} with a single bit-flip applied on bit 15 of the second binary instruction (in bold face).
The faulted \texttt{bne} instruction is decoded as a \texttt{beq}.}, label=lst:example_fault, frame=single]
loop:
  ff f2 82 93   addi t0, t0, #-1
  fe 04 [*8*]e e3   beq t0, zero, loop
\end{lstlisting}
\end{figure}

\section{Background}\label{sec:background}

\subsection{Threat Model}\label{sec:threat_model}
We consider an attacker that only has physical access to the device under attack.
The attacker is supposed to use fault injection on the device.
They can {arbitrarily} inject two kinds of faults in the memory or in the processor logic: 
either a fault with full control over a few bits (typically less than 8 bits),
or a fault altering many bits but without any control on the faulted value (random bit-flips).
They can inject multiple faults at different time locations.
{Note that state-of-the-art attackers are able to selectively inject up to 4 bit faults thanks to laser illumination~\cite{colombierMultiSpotLaserFault2022}.}
We consider fault injections targeting the instruction path only;
faults targeting the data path are assumed to be covered by a complementary dedicated mechanism
\revised{ensuring data integrity},
typically, error detection code in internal data registers and data memory.
Besides, the attacker does not have logical access to the device, and therefore cannot perform common software attacks, nor cannot modify the memory contents through logical access, e.g.\ by reprogramming it. 
Moreover, side-channel analysis and invasive attacks such as micro-probing are out of scope.

\subsection{Signature-Based Code and Control-Flow Integrity}\label{sec:signature_background}

A program can be decomposed in maximal instruction sequences with a single entry instruction and a single exit instruction, commonly called basic blocks.
A standard technique to ensure code integrity is to compute a runtime signature for each basic block from the binary encoding of its instructions~\cite{aroraHardwareAssistedRunTimeMonitoring2006, dangerCCFICacheTransparentFlexible2018}.
The signature $S_i$ associated to a basic block $\mathit{B}_i$ composed of instructions $I_0, \ldots, I_n$ is computed using a signature function $f$ and an initialization vector $\mathit{IV_i}$
\eqref{eq:signature}.
Note that fine-grained signature mechanisms are required in the context of fault injection attacks in order to detect any alteration of instructions.
Hence, the signature is usually computed from the binary encoding of every machine instruction executed.
\begin{equation}
  s_{i_0}  = f(\mathit{IV_i}, I_{0}),
  \quad
  s_{i_n} = f(s_{i_{n-1}}, I_{n}),
  \quad
  S_i = s_{i_n}
  \label{eq:signature}
\end{equation}

The runtime signature is updated each time a new instruction or sequence of instructions (e.g.\ basic block) is processed.
The runtime signature is regularly verified, for example during control-flow transfers.
Verification is usually performed by checking the signature for equality with a reference value, thereafter called \emph{reference signature}.
Reference signatures are precomputed offline, they are either stored in a dedicated memory or embedded in the program memory, e.g.\ at the end of basic blocks.

Generalized path signature analysis (GPSA) ensures a fine-grained code and control-flow integrity by computing signatures that depend on the control-flow graph~\cite{wilkenContinuousSignatureMonitoring1990}.
Typically, the signature of the basic block $\mathit{B_{i-1}}$ is used as the initialization vector $\mathit{IV_i}$ of the successor basic block $\mathit{B_i}$.
Each basic block (and each instruction in a basic block) is associated with a single and distinct signature value.
As a consequence, if several execution paths merge into a basic block, patch values 
are applied to the signature of all but one of each predecessor basic blocks $\mathit{B_j}, \mathit{B_k}, \ldots$:
an update function $u$ generates a unique initialization vector $\mathit{IV_i}$ for every tuple of signatures $S_j, S_k, \ldots$ and patch values $P_j, P_k, \ldots$
\eqref{eq:update}:
\begin{equation}
  \mathit{IV}_i = u(S_j, P_j) = u(S_k, P_k) = \ldots
  \label{eq:update}
\end{equation}

GPSA requires that reference signatures are accessible to the signature verification mechanism.
Similarly to code integrity presented above, such signatures are intertwined with program instructions, or stored in a separate memory section.
Additionally, GPSA requires to instrument the program for the application of patch values.

\subsection{Indirect Branch Integrity}\label{sec:indirect_branch_background}

Control-flow integrity (CFI) was first studied to prevent control-flow attacks on indirect branches such as ROP or JOP attacks \cite{shachamGeometryInnocentFlesh2007,bletschJumporientedProgrammingNew2011}.
The main bottleneck lies in the precise identification of the possible targets of indirect branches. As a consequence, CFI techniques rely on some over-approximations, for example \emph{equivalence classes}, to regroup targets reachable from the same indirect branch~\cite{burowControlFlowIntegrityPrecision2017}. 
Equivalence classes can be defined by various means, but usually exploit some type information associated with the target functions.
From a security perspective, the equivalence classes need to be as small as possible,
because their size define the number of targets reachable by permitted control-flow transfers~\cite{burowControlFlowIntegrityPrecision2017}.

CFI techniques typically associate a unique label to each equivalence class, and an equivalence class to each indirect branch.
The label is verified at runtime before the control flow transfer~\cite{abadiControlflowIntegrityPrinciples2009}.
Similarly, in GPSA, all the basic blocks belonging to the same equivalence class are associated to the same entry signature (i.e.\ IV value).
We call this \emph{signature confusion}.

Other techniques protect indirect branches by replacing them with sequences of direct branches~\cite{arthurGettingControlYour2015},
which removes the need for shared labels in classical CFI approaches or signature confusion in GPSA.
Note that, this approach does not prevent control-flow hijacking resulting from an alteration of the stack or of the register stroing the branch target address.

\section{\name{} Concepts}\label{sec:concept}

\subsection{\name{} Overview}\label{sec:overview}

\begin{figure}
\resizebox{\columnwidth}{!}{%
\noindent{}

\begin{tikzpicture}
    \tikzset{stages/.style={draw, inner sep=0pt, align=center, minimum height=3em, minimum width=8.4ex}}
    \tikzset{reg/.style={draw, inner sep=0pt, align=center, minimum height=4em, minimum width=2ex}}
    \tikzset{scifi/.style={draw, inner sep=0pt, align=center, minimum height=3em, minimum width=8.4ex, fill=gray!50}}
    
    \node[stages] (fetch) at (0,0) {Fetch};
    \node[reg, right=1.5ex of fetch] (ifid) {};
    \node[stages, right=1.5ex of ifid] (decode) {Decode};
    \node[reg, right=1.5ex of decode] (idex) {};
    \node[stages, right=1.5ex of idex] (execute) {Execute};
    \node[reg, right=1.5ex of execute] (exmem) {};
    \node[stages, right=1.5ex of exmem] (memory) {Memory};
    \node[reg, right=1.5ex of memory] (memwb) {};
    \node[stages, right=1.5ex of memwb] (writeback) {Write\\Back};

    \node[scifi, above=3em of execute] (signature) {CACFI};
    \node[inner sep=0, minimum size=0, left=4.25ex of signature] (a)  {};
    \node[scifi, minimum width=21.8ex, right=5ex of signature] (execintegrity) {CSI};
    \node[inner sep=0, minimum size=0, below=1em of execintegrity.208] (b) {};
    \node[inner sep=0, minimum size=0, below=1em of execintegrity.332] (c) {};


    \draw[->,>=latex] (fetch) to (ifid);
    \draw[->,>=latex] (ifid) to (decode);
    \draw[->,>=latex] (decode) to (idex);
    \draw[->,>=latex] (idex) to (execute);
    \draw[->,>=latex] (execute) to (exmem);
    \draw[->,>=latex] (exmem) to (memory);
    \draw[->,>=latex] (memory) to (memwb);
    \draw[->,>=latex] (memwb) to (writeback);

    \draw[->,>=latex] (ifid) |- (signature.160) node[near start, fill=white, align=center]{Reference\\Signatures};
    \draw[->,>=latex] (idex) |- (signature.200) node[very near start, fill=white]{Pipeline State};
    \draw[-] (exmem) |- (b);
    \draw[-] (memwb) |- (c);
    \draw[->,>=latex] (b) to (b |- execintegrity.south);
    \draw[->,>=latex] (c) to (c |- execintegrity.south);
    \draw[->,>=latex] (signature) to (execintegrity);
    \draw[-] ([xshift=-0.1ex] ifid.south east) to ([yshift=1em] ifid.south) to ([xshift=0.1ex] ifid.south west);
    \draw[-] ([xshift=-0.1ex] idex.south east) to ([yshift=1em] idex.south) to ([xshift=0.1ex] idex.south west);
    \draw[-] ([xshift=-0.1ex] exmem.south east) to ([yshift=1em] exmem.south) to ([xshift=0.1ex] exmem.south west);
    \draw[-] ([xshift=-0.1ex] memwb.south east) to ([yshift=1em] memwb.south) to ([xshift=0.1ex] memwb.south west);
\end{tikzpicture}
}
\caption{Illustration of a 5-stage processor extended with \name{} (CACFI and CSI modules, in grey)}\label{fig:overview}
\end{figure}
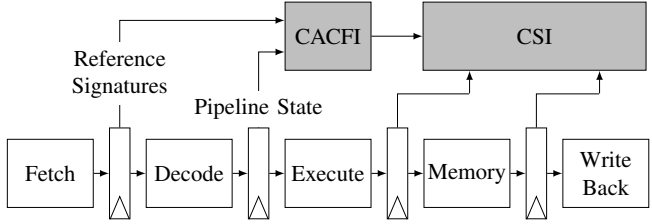

\name{} combines GPSA with a redundancy-based mechanism to ensure control-flow integrity, code authenticity, and \revised{control-signal integrity}.
Fig.~\ref{fig:overview} illustrates how \name{} would typically be integrated into a 5-stage in-order pipeline processor architecture.
\name{} is composed of two modules: 
The Code Authenticity and Control-Flow Integrity module (CACFI) implements the hardware support for GPSA and ensures \revised{control-signal integrity} up to the decode stage.
The Control Signal Integrity module (CSI) completes the coverage of \revised{control-signal integrity} through a redundancy-based mechanism.
{The two modules run in parallel with the pipeline stages and therefore do not modify the information flow within the pipeline.}
On the software side, \name{} requires modifications of the compiler backend to insert GPSA signature verifications
and patch values.

Instead of using binary encoding of program instructions to compute a signature, 
CACFI uses signals coming from the decode pipeline stage, called the \textit{pipeline state}.
CSI checks that signals from the pipeline state are correctly propagated up to their consumption in the subsequent pipeline stages.
The selected signals are duplicated into CSI at the output of the decode stage.
Then, for each subsequent pipeline stage, CSI checks the original control signals against their duplicates.
Therefore, the CSI module can detect any fault on control signals included in the pipeline state after the decode stage up to the pipeline end.
\revised{Control-signal integrity} of the whole instruction path is ensured by the combination of the CACFI and CSI modules:
CACFI ensures the integrity of the pipeline state, and CSI then ensures the integrity of control signals up to their consumption stage.

We argue that the design of a single module dealing with the control signals in all the pipeline stages,
instead of two separate modules as presented in our approach, would be increasingly more complex, if not impossible.
Indeed, many dynamic events (e.g.\ stalls due to memory latencies or jumps)
may make the computation of reference signatures and the design of the hardware module more complex.
Our decomposition into two coordinated modules avoids such complexity: the control signals selected in the decode stage are not impacted by the execution of instructions in later stages.
Moreover, it allows different implementations of the two modules as they are independent.

\subsection{Pipeline State}\label{sec:pipeline_state}
The pipeline state is a bit vector composed of control signals coming from the decode stage.
To ensure that each instruction is associated with a single signature {independently of the previously executed instructions}, each instruction must also be associated with a unique pipeline state value.
We call this property \textit{pipeline state uniqueness}.
In order to ensure code integrity, the pipeline state must include the control signals that deterministically result from the decoding of binary instructions.
Also, GPSA requires that the reference signature is computed ahead of program execution (i.e.\ by static analysis), which implies that the value of the signals monitored by the signature (and hence included in the pipeline state) can also be computed ahead of program execution. 
We discuss below which control signals can be included in the pipeline state.%

There are two kinds of control signals: the static ones and the dynamic ones.
The static control signals only depend on the instruction currently in the decode stage. 
These signals can 	be integrated in the pipeline state, since their value can be computed from the only knowledge of the related instruction.
For example, the signals for selecting the source and destination operands are directly linked to the binary encoding of instructions.
The binary encoding also contains opcode fields which control the operation to perform in the execute and memory stages.
To ensure full code integrity, the pipeline state can in addition include the contents of any immediate field in instruction encodings.

  The dynamic control signals depend on processed data or on other processed instructions.
Data-dependent control signals, such as branch decision, cannot be integrated into the pipeline state because their values cannot be statically computed.
Dynamic control signals that depend on other instructions in flight in the pipeline can be integrated to the pipeline state under certain conditions.
In the context of the processor architectures targeted by our counter-measure, that is, simple in-order processors targeting embedded systems, 
this restricts to forwarding control signals.
A forwarding mechanism enables to bypass the write-back stage when there is a data dependency between two instructions.
The computation of the forwarding signal is implementation-dependant, but without loss of generality we assume that forwarding is computed in the decode stage, and hence that its control signals can be integrated to the pipeline state.
Note that forwarding control signals that are computed after the decode stage can be protected by the CSI module.
Figure~\ref{fig:forwarding} illustrates cases where forwarding is involved. 
Figure~\ref{fig:forwarding_intra} illustrates a basic block where the forwarding is enabled between the two successive \texttt{add} instructions.
The sequence of instructions is invariable (program-dependant), the forwarding control signal can be statically determined, and hence can be safely integrated into the pipeline state.
In Figure~\ref{fig:forwarding_inter}, 
forwarding is enabled in the transition \texttt{B\textsubscript{1}} $\rightarrow$ \texttt{B\textsubscript{2}} between the \texttt{mov} and \texttt{add} instructions, 
but is disabled in the transition \texttt{B\textsubscript{2}} $\rightarrow$ \texttt{B\textsubscript{2}}  between the \texttt{bneq} and \texttt{add} instructions.
This case illustrates that forwarding may be involved at the transitions between basic blocks. 
As a consequence, the value of the forwarding control signal cannot be statically computed.
In such case, the forwarding dependency must be broken to ensure the pipeline state uniqueness property, for example by the insertion of additional instructions (Section~\ref{sec:forwarding_elimination}).
{%
  Such modification is not required when the forwarding mechanism is placed after the decode stage as its control signals cannot be included in the pipeline state.
}

\begin{figure}
  \hspace*{\fill{}}
  \subfloat[]{\begin{tikzpicture}[scale=0.5, every node/.style={scale=0.8}]
    \tikzset{bb/.style={draw, inner sep=3pt, align=left,
            minimum height=3em, minimum width=24ex,
    }}
    
    \node[bb] at (0, 0) (1) {
      \texttt{add t0, t0, \#1}\\
      \texttt{add t1, a0, t0}\\
    \texttt{load t1, 0(t1)}};
    \node[anchor=south west] at (1.north west) {\texttt{B\textsubscript{1}}};
\end{tikzpicture}\label{fig:forwarding_intra}}
  \hspace*{\fill{}}
  \subfloat[]{\begin{tikzpicture}[scale=0.5, every node/.style={scale=0.8}]
    \tikzset{bb/.style={draw, inner sep=3pt, align=left,
            minimum width=24ex,
    }}
    
    \node[bb] at (0, 0) (1) {
      \texttt{mov t0, \#0}};
    \node[anchor=south west] at (1.north west) {\texttt{B\textsubscript{1}}};

    \node[bb, below=1.5em of 1] (2) {
      \texttt{add t0, t0, \#1}\\
      \ldots\\
      \texttt{bne t0, \#16, B\textsubscript{2}}};
    \node[anchor=south west] at (2.north west) {\texttt{B\textsubscript{2}}};

    \node[inner sep=0, minimum size=0, above=.75em of 2.north east, xshift=2ex] (a) {};
    \node[inner sep=0, minimum size=0, below=.5em of 2.south east, xshift=2ex] (b) {};

    \draw[->, >=latex] (1) to (2);
    \draw[-] (2) |- (b);
    \draw[-] (b) to (a);
    \draw[->, >=latex] (a) -| (2);

\end{tikzpicture}\label{fig:forwarding_inter}}
  \hspace*{\fill{}}
  \caption{Illustration of forwarding: intra (left) and inter basic block (right)}\label{fig:forwarding}

\vspace{1em}
  \centerline{%
  \begin{tikzpicture}[scale=0.6, every node/.style={scale=0.8}]
    \tikzset{bb/.style={draw, inner sep=3pt, align=left, minimum height=3em
    				, rectangle split, rectangle split horizontal, rectangle split parts=2, rectangle split draw splits=false}}
    
    \node[bb] at (0, 0) (1) {
      \nodepart[text width=3ex]{one} \texttt{I\textsubscript{\tiny 1.0}} \\
        \ldots \\
        \texttt{I\textsubscript{\tiny 1.n}}
      \nodepart[text width=28ex, text=gray]{second} ;; $S \leftarrow S_{\scriptscriptstyle 1.0} = f(\Sigma_{\scriptscriptstyle 1.0}, IV)$ \\
        ;; $S \leftarrow \ldots$ \\
        ;; $S \leftarrow S_{\scriptscriptstyle 1.n} = f(\Sigma_{\scriptscriptstyle 1.n}, S_{\scriptscriptstyle 1.n-1})$};
    \node[anchor=south west] at (1.north west) {\texttt{B\textsubscript{1}}};

    \node[bb, below=2em of 1] (2) {
      \nodepart[text width=3ex]{one} \texttt{I\textsubscript{\tiny 3.0}} \\
        \ldots \\
        \texttt{I\textsubscript{\tiny 3.n}}
      \nodepart[text width=28ex, text=gray]{second} ;; $S \leftarrow S_{\scriptscriptstyle 3.0} = f(\Sigma_{\scriptscriptstyle 3.0}, S_{\scriptscriptstyle 1.n})$ \\
        ;; $S \leftarrow \ldots$ \\
        ;; $S \leftarrow S_{\scriptscriptstyle 3.n} = f(\Sigma_{\scriptscriptstyle 3.n}, S_{\scriptscriptstyle 3.n-1})$};
    \node[anchor=south west] at (2.north west) {\texttt{B\textsubscript{3}}};

    \node[bb, right=2ex of 2] (3) {
      \nodepart[text width=3ex]{one} \texttt{I\textsubscript{\tiny 2.0}} \\
        \ldots \\
        \texttt{I\textsubscript{\tiny 2.n}}
      \nodepart[text width=28ex, text=gray]{second} ;; $S \leftarrow S_{\scriptscriptstyle 2.0} = f(\Sigma_{\scriptscriptstyle 2.0}, S_{\scriptscriptstyle 1.n})$ \\
          ;; $S \leftarrow \ldots  $ + \textit{Patch loading} \\
          ;; $S \leftarrow S_{\scriptscriptstyle 2.n} = f(\Sigma_{\scriptscriptstyle 2.n}, S_{\scriptscriptstyle 2.n-1})$};
    \node[anchor=south west] at (3.north west) {\texttt{B\textsubscript{2}}};

    \node[bb, below=2em of 2] (4) {
      \nodepart[text width=3ex]{one} \texttt{I\textsubscript{\tiny 4.0}} \\
        \ldots \\
        \texttt{I\textsubscript{\tiny 4.n}}
      \nodepart[text width=28ex, text=gray]{second} ;; $S \leftarrow S_{\scriptscriptstyle 4.0} = f(\Sigma_{\scriptscriptstyle 4.0}, S_{\scriptscriptstyle 2.n})$ \\
        ;; $S \leftarrow \ldots$ \\
        ;; $S \leftarrow S_{\scriptscriptstyle 4.n} = f(\Sigma_{\scriptscriptstyle 4.n}, S_{\scriptscriptstyle 4.n-1})$};
    \node[anchor=south west] at (4.north west) {\texttt{B\textsubscript{4}}};

    \node[inner sep=0, minimum size=0, above=1.2em of 3] (a) {};
    \node[inner sep=0, minimum size=0, below=1em of 3] (b) {};

    \draw[->,>=latex] (1) to (2);
    \draw[->,>=latex] (2) to (4);
    \draw[-] (1) |- (a);
    \draw[->,>=latex] (a) to (3);
    \draw[-] (3) to (b);
    \draw[->,>=latex] (b) -| (4) node[near start, fill=white]{$S \leftarrow S_{\scriptscriptstyle 2.n} = u(S_{\scriptscriptstyle 3.n}, Patch)$};

\end{tikzpicture}%
  }
  \caption{Illustration of the application of GPSA to a small program sequence of 4 basic blocks.  
  The application of a patch is required in basic block $B_2$ because of the merging of two execution paths. }\label{fig:cfg}
\end{figure}

\subsection{CACFI -- Code Authenticity and Control-Flow Integrity}\label{sec:ccfi}
The CACFI module implements the hardware support for GPSA.
It requires two functions, for the signature computation and for the application of patch values, with specific properties summarized in this section.
Cf.\ Werner et al.~\cite{wernerProtectingControlFlow2015} for a detailed discussion.
Note that most cryptographic functions intrinsically support all these properties.

\subsubsection{The signature function}\label{sec:signature}
{%
The signature function $f$ is the core of GPSA.
In CACFI, $f$ computes the runtime signature from the pipeline state and the previous runtime signature.
The runtime signature is stored in the \textit{signature register} within CACFI.
The signature register should not be directly accessible from any instruction to limit the attack surface on CACFI.
The GPSA fault detection capabilities depend on $f$'s properties.
}
\begin{enumerate}
  \item \emph{Collision resistance}: prevents an attacker from forging a faulted basic block presenting the same signature as the signature of the original basic block (also known as second-preimage resistance). 
  This property also prevents the attacker from reverting the signature to a valid value after the introduction of one or many faults.
  \item \emph{Error preservation}: alterations of signature values are not cancelled by any following fault-free sequence.
This property, in combination with collision resistance, allows for the arbitrary placement of signature verifications.
  \item \emph{Non associativity}: sequences of instructions with different orderings produce different signatures.
This property ensures control-flow integrity at the granularity of machine instructions.
  \item \emph{Invertibility}: is introduced by \cite{wernerProtectingControlFlow2015} to compute patch values.
  However, it is not required by our approach because patch values are applied on the basic block signature instead of being applied on intermediate signatures (see below).
\end{enumerate}

Note that many function signatures can support the properties listed above, which allows for many implementation trade-offs.
For example, 
key-based cryptographic signature functions such as Message Authentication Codes (MAC) are good candidates.
MACs require a secret key, which prevents the generation of valid reference signatures without knowledge of the secret key.
Using such function signatures, \name{} ensures code authenticity in addition to code integrity.
In Section~\ref{sec:implementation}, we present two implementations, one using CBC-MAC with the Prince block cipher and another one, only supporting code integrity and not code authenticity, using an error detection code.

\subsubsection{The update function}\label{sec:update}
GPSA requires the application of an update function $u$ before merging execution paths.
This function must support the following properties:
\begin{enumerate}
  \item \emph{Full control}: given a signature, there exists a patch value for any target IV.
  \item \emph{Error preservation}: any fault previously introduced in the signature cannot be reverted by applying an error-free update.
  \item \emph{Invertibility}: a patch value can be computed from an initialization vector and a signature.
\end{enumerate}
 
The update function is triggered at each control-flow transfer (i.e.\ taken branch, call and return).
The runtime signature in the {signature register} is updated using function $u$ and the current patch value.
The patch value is stored in a \textit{patch register} in CACFI,
and can be updated by a dedicated instruction that loads a patch value from memory.
Additionally, the patch register is reset to a default, constant patch value after the processing of each control-flow instruction (taken or not).
This default patch value must be known at compile time to compute the reference signatures, and the identity element of $u$, if it exists, can be used as the default patch value.

When several basic blocks $\mathit{B_i},\mathit{B_k},\ldots$ have the same successor $\mathit{B_s}$, there is at most a single basic block $\mathit{B_f}$ falling into $\mathit{B_s}$ (i.e.\ the basic block immediately preceding  $\mathit{B_s}$ in the memory layout).
If $\mathit{B_f}$ exists, its signature $\mathit{S_f}$ is used as the  initialization vector $\mathit{IV_s}$ of $\mathit{B_s}$: $\mathit{IV_s}=\mathit{S_f}$.
Otherwise, $\mathit{IV_s}$ is chosen randomly among the signatures of $\mathit{B_i},\mathit{B_k},\ldots$.
Knowing $\mathit{IV_s}$ and $u^{-1}$, a patch value is computed for all the other  predecessors of $\mathit{B_s}$.
Fig.~\ref{fig:cfg} shows a simple example of a CFG that requires an update on one of its edges.
The instructions in basic blocks \texttt{B\textsubscript{1}}, \texttt{B\textsubscript{2}} and \texttt{B\textsubscript{3}} do not require an update because they all have only one predecessor.
The instruction \texttt{I\textsubscript{4.0}} has two predecessors, \texttt{I\textsubscript{2.n}} and \texttt{I\textsubscript{3.n}}, and therefore requires an update.
\texttt{B\textsubscript{3}} falls into \texttt{B\textsubscript{4}} which means that if \texttt{I\textsubscript{3.n}} is a branch, then it is not taken on this execution path.
Therefore, it is not possible to have an update on the execution path \texttt{B\textsubscript{3}}--\texttt{B\textsubscript{4}}.
This is why the update is applied in \texttt{B\textsubscript{2}} during the control transfer to the taken branch, i.e.\ \texttt{B\textsubscript{4}}.

\subsubsection{Signature verification}\label{sec:verification}
A runtime signature is computed for every instruction in the program.
Thanks to the properties of the functions $f$ and $u$, any fault captured in the signature will be forwarded into the next ones (cf.\ \ref{sec:signature}).
Therefore, it is possible to insert verifications anywhere in the program.

\name{} uses custom control-flow transfer instructions, thereafter called \emph{verification instructions}, which have the same semantics as their original counterpart.  
Verification instructions load a reference signature immediately following in the program memory, and trigger the signature verification.
Then, they proceed similarly to other control-flow instructions: if the branch is taken, the runtime signature is updated with the current patch value.
Finally, the current patch value is reset to its default value.

The substitution of  a control-flow instruction by a verification instruction impacts code size, as a reference signature is inserted after each verification instruction, and potentially execution time if the delay due to the loading of the reference signature is not masked.
When the verification fails, it triggers an exception that calls a software user-designed fault handler.

Thanks to the use of verification instructions as control-flow instructions, our approach provides great flexibility in the insertion of signature verifications,
which allows to fine-tune the trade-off between the detection delay and the overheads due to code size and execution time.
Similarly to GPSA, it is possible to use a single verification instruction at the exit point of a secured function to minimize the performance overheads without reducing the detection coverage of the counter-measure.
We discuss the security impact of such trade-offs in Section~\ref{sec:security_analysis}.

\subsection{CSI -- Control Signal Integrity}\label{sec:csi}

The CSI module ensures \revised{control-signal integrity} for the pipeline stages following the decode stage.
The principle is to use a redundancy scheme to detect any change in the control signals constituting the pipeline state, from their emission to their consumption stage.
This approach is lightweight because it involves only a small part of the pipeline's control logic.
The CSI module duplicates the propagation of selected signals between the different stages in the pipeline.
In each pipeline stage, the duplicated signals are checked against the original ones.
The duplication can use any redundancy scheme, potentially with several duplicates, e.g.\ a simple copy, a complementary copy or the initial value \code{xor}ed with an arbitrary value.

\subsection{Indirect Control-Flow Handling}\label{sec:indirect_control_flow}

In this section we focus on the protection of indirect function calls and function returns.
Other indirect branches can be removed using a compiler option.
We discuss this in Section~\ref{sec:software_support}.

\name{} uses equivalence classes derived from function prototypes to identify indirect call targets.
Equivalence classes regroup functions with identical function prototypes (return type, number of arguments and type of each argument).

\name{} combines GPSA with indirect call elimination to remove signature confusion, hence reducing the attack surface.
Each indirect function call is replaced by a \emph{dispatcher} (illustrated in Listing~\ref{fig:dispatcher}), that is, a sequence of direct branch instructions that forwards the control flow to the target function.
With this approach, each function remains associated with a unique IV even if the function is the target of indirect branches.

For function returns, which are also indirect branches, \name{} uses signature confusion only, although it would be possible to use dispatchers.
All the basic blocks that follow the calls to a given function belong to the same equivalence class, and hence share the same IV.
When a function has several exit points, \name{} assumes a constant signature value at each of the exit points.
As a consequence, patch values are applied to all but one of the exit points.
Note that without indirect call elimination, all the function sharing the same indirect call site would also share the same signature at their exits, hence increasing the attack surface.
To increase the protection level of function returns, it is possible to extend \name{} with a shadow stack.

\begin{figure}
  \centering
\begin{lstlisting}[captionpos=b,caption={Example of indirect function call in language C}, label=fig:indirect, frame=single, language=C]
void bar();
void baz();

void foo(void (*fptr)()) {
  /* fptr is either assigned to &bar or to &baz */
  fptr();
}
\end{lstlisting}

  \centering
\begin{lstlisting}[captionpos=b, caption={Protection of the source-code example from Figure~\ref{fig:indirect} with \name{} (in RISC-V pseudo assembly).   The indirect function call is replaced by a dispatcher.   In the original code of function \code{foo}, register \code{a0} stores the branch target address.  }, label=fig:dispatcher, frame=single, language={[riscv]Assembler}]
foo:
  call        dispatcher_EC0_a0_0
  ret

dispatcher_EC0_a0_0:
  push        ra
  push        s11
dispatcher_EC0_a0_0_bar:
  li          s11, bar
  bne         s11, a0, dispatcher_EC0_a0_0_baz
  load_patch  PATCH_bar
  call        bar
  load_patch  PATCH_ret_dispatcher
  jmp         dispatcher_EC0_a0_0_ret
dispatcher_EC0_a0_0_baz:
  li          s11, baz
  bne         s11, a0, error_handler
  load_patch  PATCH_target1
  call        baz
dispatcher_EC0_a0_0_ret:
  pop         ra
  pop         s11
  ret
\end{lstlisting}
\end{figure}

\subsection{Branch Prediction}\label{sec:branch_prediction}
{%
When the pipeline implements branch prediction, the control flow might roll back to the previous branch in case of a misprediction.
The pipeline then flushes the speculatively executed instructions and the execution resumes at the correct address.
In order to be compatible with branch prediction, 
CACFI saves the signature register after a branch in order to support signature roll back.
On misprediction, the signature register is restored to the saved signature.
When a branch is predicted not taken, the update function is applied to the signature register before saving it so that in case of misprediction the restored signature is the one that would have been computed if the branch had been taken.
}

After a misprediction, the instructions in the pipeline are invalidated but may impact the value of the dynamic control signals of the next pipeline state,
which may impact the pipeline state uniqueness (Section~\ref{sec:pipeline_state}).
Since \name{} already requires breaking dependencies such as forwarding dependency at basic block transitions, 
branch prediction does not add more constraints to the CACFI module.
In Section~\ref{sec:forwarding_elimination}, we propose to break forwarding dependency at basic block boundaries using a dedicated compiler pass.

To protect speculatively executed instructions and invalidated instructions, the CSI module should also cover all the control signals related to branch prediction.

In conclusion, branch prediction can be supported by \name{} with a negligible increase of the complexity of the CACFI and CSI modules. 

Note that the branch prediction mechanism itself remains sensitive to some fault injection attacks.
\name{} is not able to detect a fault targeting the misprediction control signal that would change the branch decision.
To protect against such case, it would be necessary to ensure data integrity or/and to duplicate the misprediction control signal.
Any other fault affecting the control flow is detected by \name{}.

\subsection{Interrupts Handling and Protection}\label{sec:interruptions}

Interrupts can occur at any time during program execution and hence interrupt handlers cannot be associated with a set of predecessor instructions.
As a consequence, a dedicated mechanism is required to handle interrupts and to protect interrupt handlers.
\name{} is designed to fully protect the execution of interrupt handlers and to increase the difficulty to leverage interrupts in an attack scenario.

Each interrupt handler is associated to a different IV, and all the IVs are stored in a table similar to the interrupt vector table.
Upon triggering of an interrupt, the {signature register} is saved in a dedicated register, called the \emph{context register}.
The CACFI module selects the IV corresponding to the triggered interrupt to reset the {signature register},
and the processor starts the execution of the interrupt handler.
A verification instruction can be placed at the end of the interrupt handler to ensure its integrity.
Similarly to other sequences of code, verification instructions can be added inside the interrupt handler if needed to reduce the delay between verifications.
When the interrupt handler returns, the {signature register} is restored to the value saved in the context register.

After the interrupt handler has returned, the last instructions of the interrupt handler are still in the pipeline.
This might impact the pipeline state uniqueness the same way as forwarding dependency between basic blocks (Section~\ref{sec:pipeline_state}).
To avoid this, 
\name{} delays interrupt processing until the end of a basic block, 
since forwarding dependencies are already broken at basic block transitions.

In our design, the signature register is not saved in memory, which prevents attacks on the saved signature, e.g.\ during memory transactions.
This helps reduce the attack surface of interrupts, for a negligible hardware overhead, and allows for the use of dedicated protections on the context register if need be.
To support nested interrupts, a \emph{context stack}, internal to the processor, can be used in place of the context register.

\section{Implementation}\label{sec:implementation}
We integrate \name{} to the CV32E40P
processor~\cite{openhwgroupCV32E40P2021},
a 32-bit, in-order,  4-stage \mbox{RISC-V} core implementing the RV32I base instruction set version 2.1.
We select the CV32E40P because such a small in-order core is representative of typical fault injection targets,
and because a 4-stage pipeline is representative of the main challenges of microarchitectural design due to control and data hazards such as forwarding mechanisms.
The integration to more complex processors is left for future work.

\subsection{Pipeline State}

We manually select the control signals to integrate to the pipeline state.
The pipeline state consists of 64 bits composed as follows:
\begin{enumerate}
  \item All the non-redundant control signals internal to the decode stage that are involved in the operand selection and the forwarding mechanisms:
      23 bits from the operand selection multiplexers;
      4 bits from the operand forwarding multiplexers.
  \item All the control signals produced by the decode stage and transmitted to the next stages and that deterministically result from the decoding of the instruction opcode:
7 bits to control the arithmetic and logic unit;
      2 bits to control the read and write enable of the load store unit;
      10 bits to control the registers to write in the write-back stage.
  \item All the signals derived from the immediate data fields:
      10 bits from the binary instruction's immediate fields.
        Note that RISC-V ISA supports up to 20 bits immediate, but the remaining immediate fields overlap with the operand selection fields that are already included in the pipeline sate.
  \item 8 bits of padding to fill the 64 bits of the pipeline state.
\end{enumerate}

The control signals outputted by the decode stage and that will go through subsequent stages are duplicated in the CSI module.
The remaining ones are directly used after the decode stage and do not go through more than one other stage.
We use a simple duplication scheme to implement the CSI redundancy.

\subsection{Signature and Update Functions}

\name{} is implemented with two different single cycle signature functions for the CACFI module:
\begin{itemize}
  \item a CBC-MAC based on a fully unrolled hardware implementation of the Prince block cipher, which is selected for its small silicon area 
  \revised{and for its capability to deliver output within one CPU clock cycle}.
    Prince is a symmetric cipher using 64-bit blocks and a 128-bit key, and the CBC-MAC therefore generates 64-bit tags.
    In order to limit the code size and runtime overheads, 
    the CACFI signature is composed of the 32 lowest significant bits of the CBC-MAC output tag.
    We discuss the security impact of this design choice in Section~\ref{sec:security_analysis}.
  \item a CRC32 designed to detect up to 8 bit-flips per basic block.
    Because CRC32 functions do not use any secret to compute the signature, \name{} only ensures code integrity and not code authenticity in this case.
\end{itemize}
We select the exclusive or (XOR) for the update function in the two implementations.

\subsection{\name{} RISC-V ISA Extension}
The CV32E40P is modified as follows.
All the control-flow instructions update the runtime signature with the current patch value if the branch is taken (Section~\ref{sec:update}).
The core is also extended with custom instructions:
we implement a verification instruction (Section~\ref{sec:verification}) for each control-flow instruction in the RV32I instruction set
(\code{\name{}.beq}, \code{\name{}.bne}, \code{\name{}.blt}, \code{\name{}.bge}, \code{\name{}.bltu}, \code{\name{}.bgeu}, \code{\name{}.jal}, \code{\name{}.jalr}).
We also add a load patch instruction (\code{\name{}.ldp}) that fetches a patch value from memory, in the \code{.patches} section;
the base address is stored in a new Control Status Register (CSR).
This CSR is set during the core bootstrap to point to the memory section of the binary program that gathers all the patches.
The patch value offset in the memory section is encoded as a 20-bit immediate value in the \code{\name{}.ldp} instruction,
which limits the number of patch values accessible to $2^{20}$.
If patch values are aligned on 4-byte boundaries in memory, it is possible to fix to 0 the two least significant bits of the offset, and increase the number of accessible patch values to $2^{22}$.

Note that it is not desirable to store patch values in the immediate fields of instruction encoding, because this would introduce a circular dependency between the signature values and the patch values.
A solution to avoid this dependency is to remove the immediate fields from the pipeline state for \code{\name{}.ldp} instructions,  hence increasing the complexity of the decoding logic.
Furthermore, locating patch values in a dedicated section offers the possibility to store the values in a separate, secured memory.

\subsection{Software Support}\label{sec:software_support}
\begin{figure}
\includegraphics[width=\linewidth]{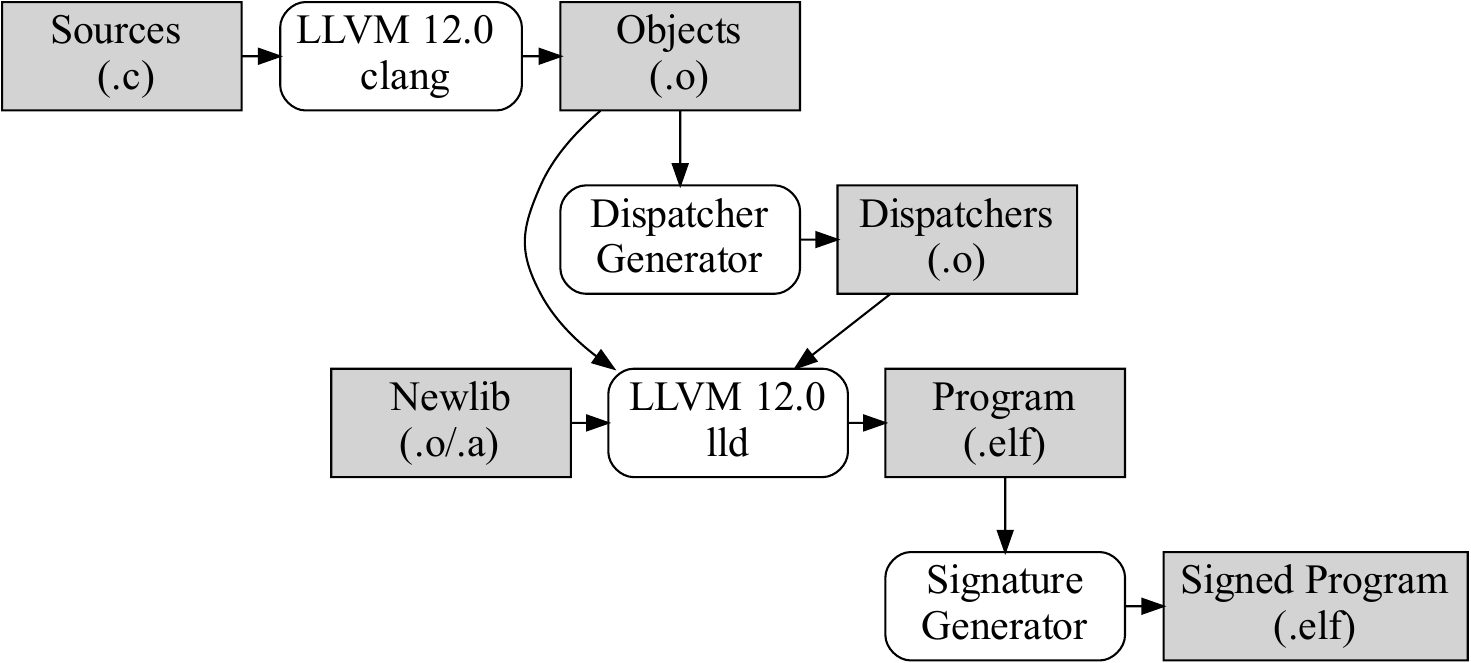}
  \caption{\name{} compiler toolchain with the files in light gray and the tools in rounded boxes}\label{fig:toolchain}
\end{figure}

Fig.~\ref{fig:toolchain} presents the complete \name{} compilation process.
We select LLVM version 12.0 to build \name{} compiler toolchain.
We extend the RISC-V backend with several passes to apply the code modification required by \name{}.

\subsubsection{Forwarding dependency elimination pass}\label{sec:forwarding_elimination}
This pass ensures the pipeline state uniqueness property at basic block transitions (Section~\ref{sec:pipeline_state}), which is mandatory to support basic blocks with multiple predecessors, branch prediction and interrupts.
For the first instructions of a basic block, this pass ensures that the forwarding mechanism is deactivated for all the predecessors and if need be inserts a \code{nop} instruction to break the forwarding dependency.
The insertion of a single instruction is sufficient for the 4-stage CV32E40P core, but longer instruction sequences may be required for more complex pipeline architectures.%

\subsubsection{Dispatcher pass}
This pass replaces each indirect call by a direct call to a dispatcher (Section~\ref{sec:indirect_control_flow}).
The code for each dispatcher is generated separately by the Dispatcher Generator before the linking process.

Indirect branches that do not represent function calls are eliminated using the \code{-fno-jump-table} option.

\subsubsection{Patch placement pass}
This pass performs a control-flow analysis to insert the \code{\name{}.ldp} instructions.
As described in Section~\ref{sec:update}, \name{} requires a patch for all but one predecessor for each basic block.
A \code{\name{}.ldp} is also inserted before each call instruction and all but one functions returns.
Loops require a dedicated analysis regarding the update function to prevent circular dependencies during the reference signature computation.
The simple rule of placing a \code{\name{}.ldp} in all but one predecessor can fail to break such a circular dependency.
In this case, the pass adds a \code{\name{}.ldp} in one of the loop's basic blocks.
Note that this pass disables tail call optimization only when a function has multiple exit points so that the caller function and the tail-called function do not share the same signature.

The patch values and offsets in the \code{.patches} section are computed later by the Signature Generator tool.

\subsubsection{Reference signature placement pass}
The third pass identifies all the functions annotated with the dedicated attribute, \code{\name{}\_secured}, and replaces the control-flow instructions by the \name{} equivalent ones that trigger the signature verification.
This pass inserts a signature placeholder after each branch that performs a signature verification.

\subsubsection{Dispatcher Generator}
Before the link process, this tool builds the dispatchers by leveraging debug information from the \code{clang} compiler.
The Dispatcher Generator computes the equivalence classes through a context-insensitive analysis from the type information of the target function prototypes (Section~\ref{sec:indirect_control_flow}).
Each equivalence class is associated with one dispatcher for each register storing a target address in an original function call (e.g.\ register \code{a0} in Figure~\ref{fig:dispatcher}).
From the perspective of an attacker, this design choice increases the difficulty to fault the target function address because the address can be stored in different registers, depending on the register allocation policy used by the compiler.

\subsubsection{Instrumentation of the Newlib C-library}
We use LLVM infrastructure to link the sources object files with the dispatchers and the Newlib C-library.
The C-library is not protected with verification instruction, but it is instrumented with signature updates (\code{\name{}.ldp} instructions) so that the signatures are correctly propagated through the library.
Yet, extending the C-library with signature verifications only requires minimum changes by adding the \code{\name{}\_secured} attribute to the desired function prototypes.

\subsubsection{Signature Generator}
Post link, the Signature Generator extracts the CFG by static analysis.
Then, it computes the reference signatures and patch values.
The computation is done by exploring the CFG recursively basic block per basic block.
Each basic block is processed through a stateful signal-accurate model of the CV32E40P's decode stage to extract the pipeline state and derive the signature.
When a \code{\name{}.ldp} is present in a basic block, the Signature Generator attributes it a unique offset in the \code{.patches} section.
The associated patch value is computed from the basic block signature and its successor signature.
Finally, the Signature Generator creates a new ELF file with the reference signature placeholder and the \code{\name{}.ldp} offset filled and the additional \code{.patches} section containing the patch values.

\section{Security Analysis}\label{sec:security_analysis}

This section presents a security analysis of \name{} regarding our threat model.
\name{} is designed to ensure code authenticity or code integrity, control-flow integrity and \revised{control-signal integrity}.
The data integrity property is not supported by \name{} and is supposed to be ensured by a complementary dedicated mechanism.

\subsection{Pipeline State Verification}

\begin{revisedenv}
The pipeline state is the cornerstone of \name{} as the control signals included in the pipeline state feed both the CACFI and CSI modules.
The construction of the pipeline state determines the capability of \name{} to ensure code integrity and \revised{control-signal integrity}.
As presented in Section~\ref{sec:implementation}, in our implementation, we build the pipeline state through a manual analysis of the control signals emitted by the decode stage.
Such analysis and the resulting implementation are prone to errors, which could lead to vulnerabilities.

\subsubsection{Verification Workflow}

We perform a formal verification of our implementation of the pipeline state  and of the \revised{control-signal integrity} property 
with the workflow of Tollec et al.~\cite{tollecExplorationFaultEffects2022}, which targets the formal vulnerability analysis against fault injection attacks from both a hardware (RTL) and a software (binary code) description. 
The workflow works as follows.
The processor implementation, in SystemVerilog, is translated using the Yosys tool~\cite{YosysyosysYosysOpen2023} into a formal model in the Satisfiability Modulo Theory Language 
(SMT-LIB), which represents at the RTL level the combinatorial and sequential logic of the CPU, memories, and peripherals.
This translation preserves a complete correspondence between RTL signals and SMT variables.
The binary program 
is expressed, in the Yosys SMTC constraints language, 
as constraints applied to the SMT model of the processor memories (typically, the RAM). 
The fault model, specified by the user in SMTC, defines: 
the target SMT variables, 
the effect of fault injection, 
the maximum number of injections, 
and the timing constraints (CPU cycles where fault injection is possible).
The fault model is automatically instantiated as controllers applied to all the target SMT variables.
The verification engine performs Bounded Model Checking (BMC), leveraging the Yosys-SMTBMC tool for BMC and Yices\,2 for satisfiability queries.
During verification, the BMC engine drives the fault controllers, 
and verification targets a property $\varphi$ specified by the user.
When $\neg \varphi$ is satisfiable (i.e.\ $\varphi$ does not hold), the workflow generates a VCD trace  as a counter-example. 

 \subsubsection{Verification Use Case and Verified Properties}
 
We verify the security of MAFIA running a VerifyPIN program under fault injection.
VerifyPIN is an authentication procedure where an input (user) PIN code is compared to a secret (card) code.
This small program includes control flow (conditional and unconditional branches, function calls), memory accesses, etc.,  and is similar to the \code{memcmp} procedure widely used in software.
Hence, it is representative of the many ways to leverage fault injection in an attack of the authentication procedure, e.g.\ bypassing authentication, bypassing the PIN comparison, modifying the status value returned, altering the computation of PIN comparison.
Data integrity is not verified (although it can be) because it is not included in our threat model.
The formal verification targets 
$    \neg \varphi := 
    	\psi{}
    	\land 
 	\neg \phi{}_1 
 	\land
 	\neg \phi{}_2 
$, where properties $\psi{}$, $\phi{}_i$ are informally described as follows:
\begin{description}

\item[$\psi{}$] Authentication succeeds with different user and card PIN codes.

\item[$\phi{}_1$] Fault injection leads to an alteration of the pipeline state. 

\item[$\phi{}_2$] Fault injection is detected by the CSI module.
\end{description}
If $\phi{}_1$ holds, we know that the CACFI module ensures code and control-signal integrity, 
since any alteration of the pipeline state is captured by the signature function (Section~\ref{sec: security analysis: signature functions}).
If $\phi{}_2$ holds, we know that the CSI module ensures control-signal integrity for the pipeline stages after decode.

The verification considers a single fault (mono- or multi-bit) applied to any control signal of size smaller or equal to 8 bits, at each of the CPU cycles in a 60-cycles window overlapping with the full execution of the VerifyPIN procedure.
Note that, in our target implementation, the control signals of size larger than 8~bits don't need to be considered since they drive unused features, e.g.\ multiplier, performance monitoring module.

\subsubsection{Verification Results}

The verification fails to find any fault leading to an invalidation of property $\varphi$.
Furthermore, \name{} eliminates all the vulnerabilities identified by the same workflow on the original (unprotected) CV32E40P processor running the VerifyPIN procedure.
Note that, due to the error preservation property of the signature function, multiple fault injections that do not exceed the attacker capabilities in our threat model (e.g.\ 8 cumulative bit flips for the CRC32 implementation) are also detected by the CACFI module.  
Multiple fault injections may not be captured by our implementation of the CSI module, but can be supported by other redundancy schemes for a negligible increase in overhead (Section~\ref{sec:csi}).

This brings confidence in the pipeline state construction and in \name{}'s capacity to protect the execution of an application against fault injections in the microarchitecture.

\end{revisedenv}

\subsection{Signature Functions}
\label{sec: security analysis: signature functions}

The CBC-MAC/Prince signature function uses a secret key, which prevents an attacker from inverting the signature function to identify collision values.
Furthermore, CBC-MAC with Prince provides strong resistance to collision attacks because in our threat model the attacker cannot control the whole contents of a basic block.
In our implementation, only 32 bits of the 64-bit signature are verified, which reduces the probability to find a collision to $1/2^{16}$ because of the birthday paradox.
Yet, a successful collision attack is unlikely because the complexity of this attack combines with the complexity of fault injection.
Last, the known weaknesses of CBC-MAC, such as message forgery for variable length messages or variable initialization vectors, are not relevant here because our threat model assumes that it is not possible to modify the memory contents except by the use of fault injection.

The CRC32 signature function protects against a weaker attacker model, because it is designed to ensure code and control-flow integrity only.
{%
Moreover, CRC functions are invertible, meaning that an attacker could  identify the fault to inject to create a signature collision.
Yet, CRC functions are designed so that collisions are possible only above a fixed amount of bit-flips.
Therefore, instead of relying on direct collision resistance, CRC functions rely on the detection capabilities to increase the fault injection complexity.%
}

For CRC32, We determine the best candidate function according to our security model.
To do so, we search, in a list of polynomials known for providing good detection capabilities~\cite{koopman32bitCyclicRedundancy2002}, the polynomial that requires the highest minimal number of bit-flips to create a collision in the signature.
The search tests exhaustively all the basic block lengths up to 40 instructions, and all the collision vectors with a Hamming Weight value smaller than 11.
The best generator polynomial identified is \code{0xFA567D89}, which detects up to 8 bit-flips in the input sequence.
Thus, in order to create a collision in the signature, an attacker has to control precisely the alteration of at least 8 bits in the pipeline state.
Note that such collision can be obtained in one or several fault injections.
It can also be obtained indirectly by faults targeting the update function (including the patch value) or the runtime signature value, since these values are in the end combined with subsequent values of the pipeline state.
However, such fault targets do not reduce the security level of the candidate CRC32 function.

\subsection{Signature Verification}
A possible attack is the case where a fault triggers a jump outside the program sections instrumented with signature verifications.
This attack is equivalent to the case where several faults target all the subsequent verifications after a first fault.
In such case, the attack is undetected by the counter-measure, and the security level of \name{} is determined by the time intervals between verifications.
As the substitution of control-flow instructions by verification instructions is performed at compile time, it is possible to determine the maximum delay between successive verifications, or to constrain it by inserting extra verification instructions (i.e.\ direct branch instructions jumping to the instruction following in program memory).
A watchdog could then ensure that this maximum delay is never reached, {and so detect an attacker jumping outside the program section instrumented with signature verification.}

\subsection{Control Signal Integrity}

The CSI module covers all the control signals transmitted from the decode stage to the next stages.
An attack targeting the pipeline stages after the decode stage can be effective if it simultaneously faults the original signal and its duplicate in the CSI module.
If such an attack is relevant, it can be mitigated by using redundancy schemes with better detection capabilities,
and we believe that the implementation of such schemes will have a negligible impact on the hardware area overhead, since the number of control signals monitored by CSI is low.
Additionally, the comparison result is encoded as a single bit connected to the exception mechanism.
A single fault could then prevent the detection propagation and the software handler triggering.
A typical protection is the use of specific encoding (e.g.\ differential encoding) for the connection to the exception mechanism.

\subsection{Control-Flow Integrity}

\name{} 
ensures a static CFI policy, which ensures that:
i)~for indirect branches, the target address is part of the identified equivalence class;
ii)~for returns, the target address follows a valid call site of the current function for return.
\name{} thus considerably reduces the number of reachable addresses that would not be detected.
Thanks to the combination 
of CFI
with code authenticity and \revised{control-signal integrity}, the replacement of a control-flow target address by a valid address is restricted to data corruption, which is outside our threat model.

In the following, we proceed with an in-depth security analysis of our design, considering the possibility of attacks outside our threat model.
Regarding the protection of indirect branches,
the security level is impacted by the precision of the analysis of indirect branch targets, e.g.\ the size of equivalence classes (Section~\ref{sec:indirect_branch_background}).
Table~\ref{table:dispatchers} reports an analysis of the evaluated benches with indirect branches (Section~\ref{sec:experimental_evaluation}), 
showing the number of dispatchers inserted, the number of equivalence classes, the size of each equivalence class, and the number of non-legitimate functions per class.
The number of dispatchers depends on the number of indirect call sites in the original bench.
The classes size reports the number of reachable functions in each equivalence class.
In the evaluated benches, the number of equivalence classes does not exceed 2 and the largest class is limited to 9 elements, which fairly restricts the possibility of attacks.
Note that several dispatchers can correspond to the same equivalence class (but using different registers for the target branch address): for \code{wikisort} and for \code{sglib-combined -Os} there are more dispatchers than equivalence classes.
{%
Finally, non-legitimate functions are functions that are not reachable from a given call site.
In an equivalence class, the non-legitimate functions correspond to functions unreachable from an indirect call site but that share the same property (e.g.\ prototype) with the other functions.}
\begin{revisedenv}
Out of 3 benches, \code{sglib-combined} is the only one to present equivalence classes
that include non-legitimate functions, 
and furthermore each class is composed of non-legitimate functions only.
Here, the context-insensitive analysis performed by the Dispatcher Generator is not able to detect that there is a function pointer set to \code{NULL} in the source code, 
resulting in one or two useless dispatchers depending on the optimization level.
Such issue could be avoided by manually selecting the dispatchers to include into the program after a cross-analysis of the source code and of the metrics reported by the Dispatcher Generator.
In conclusion, even against an attacker able to control data, \name{} only leaves a small attack surface against attack such as ROP or JOP. 
\end{revisedenv}

\begin{table}[]
  \centering
  \caption{\revised{%
  Analysis of the benches including indirect branches: number of dispatcher functions, number and sizes of classes,  and number of non-legitimate functions in each class.%
  }%
  }\label{table:dispatchers}
\begin{tabular}{|l|r|r|r|r|}
\hline
Bench
	& \revised{\begin{tabular}{@{}c@{}}Nb.\ dis-\\patchers	\end{tabular}}
	& \revised{\begin{tabular}{@{}c@{}}Nb.\ eq.\\classes	\end{tabular}}
	& \revised{\begin{tabular}{@{}c@{}}Classes\\size		\end{tabular}}
	& \revised{\begin{tabular}{@{}c@{}}Nb.\ non-leg.\\fun.\ in classes	\end{tabular}}
	\\
\hline
picojpeg (\code{Os})        & 1 & 1 & [1]     & [0]\\
\hline
picojpeg (\code{O2})        & 2 & 1 & [1]     & [0]\\
\hline
sglib-combined (\code{Os})  & 2 & 2 & [1, 3]  & [1, 3]\\
\hline
sglib-combined (\code{O2})  & 1 & 1 & [3]     & [3]\\
\hline
wikisort (\code{Os})        & 9 & 2 & [1, 9]  & [0, 0]\\
\hline
wikisort (\code{O2})        & 3 & 2 & [1, 9]  & [0, 0]\\
\hline
\end{tabular}
\end{table}

\section{Experimental Evaluation}\label{sec:experimental_evaluation}

To evaluate the hardware overhead due to \name{}, we synthesize the modified CV32E40P into an Application Specific Integrated Circuit (ASIC).
The ASIC is designed for a frequency of 400MHz, in the GF-22FDX FDSOI technology, and the target frequency is not impacted by the addition of \name{}.
The core occupies
	64~kGE with CBC-MAC/Prince and
	55~kGE with CRC32,
which represents an area overhead wrt.\ the unmodified core of
	23.8\% and
	6.5\% respectively.

The software evaluation is carried out through HDL cycle-accurate simulations of the modified CV32E40P with CRC32.
We benchmark our implementation with the Embench-IoT~\cite{EmbenchOpenBenchmarks2021} test suite, which targets embedded systems without operating system.
All the test programs are compiled with the \name{} toolchain, with optimization levels \code{-Os} and \code{-O2}, and are linked with the Newlib C-library and the LLVM multiplication and soft float libraries.
The benches are compiled with the option \code{-ffunction-section} to eliminate any dead code.
Embench-IOT contains 4 benches with indirect branches.
We prevent the compiler from using indirect branches with the \code{-fno-jump-table} compiler option, which leaves 3 benches with indirect function calls (\code{picojpeg}, \code{sglib-combined} and \code{wikisort}).
Note that \code{slre} could not be compiled with the \code{-O2} optimization level because our reference signature generation does not support the inter-procedural loop caused by a recursive call.

We add the attribute \code{\name{}\_secure} to the benchmarked functions only, meaning that only those functions contain signature verifications.
The C-library and the LLVM library do not contain any signature verification but are still instrumented with \code{\name{}.ldp} instructions.

The code size evaluation considers only the sections impacted by \name{} (\code{.text} and \code{.patches}), which provides a pessimistic, upper bound of the overall code size overheads for a complete firmware image.

\begin{table*}[]
  \caption{Embench-IOT result with the size (in bytes, and overhead wrt. unprotected version), signatures and patches and the execution time (in CPU cycles, and overhead wrt. unprotected version)  for \name{} with the CRC signature function}\label{table:result}
\centering
\begin{tabular}{|l|rrrr|rrrr|}
\hline
\multirow{2}{*}{Bench} & \multicolumn{4}{c|}{O2} & \multicolumn{4}{c|}{Os} \\ \cline{2-9} 
& \multicolumn{1}{r|}{Size} & \multicolumn{1}{r|}{Signatures} & \multicolumn{1}{r|}{Patches} & Exec. time & \multicolumn{1}{r|}{Size} & \multicolumn{1}{r|}{Signatures} & \multicolumn{1}{r|}{Patches} & Exec. time \\ \hline
aha-mont64 & \multicolumn{1}{r|}{6204 (\texttimes 1.30)}  & \multicolumn{1}{r|}{124}  & \multicolumn{1}{r|}{106}  & \multicolumn{1}{r|}{75335 (\texttimes 1.38)}  & \multicolumn{1}{r|}{4720 (\texttimes 1.28)}  & \multicolumn{1}{r|}{92}  & \multicolumn{1}{r|}{70}  & \multicolumn{1}{r|}{67461 (\texttimes 1.18)} \\ \hline
crc32 & \multicolumn{1}{r|}{824 (\texttimes 1.34)}  & \multicolumn{1}{r|}{8}  & \multicolumn{1}{r|}{21}  & \multicolumn{1}{r|}{380997 (\texttimes 1.21)}  & \multicolumn{1}{r|}{856 (\texttimes 1.35)}  & \multicolumn{1}{r|}{9}  & \multicolumn{1}{r|}{21}  & \multicolumn{1}{r|}{381006 (\texttimes 1.21)} \\ \hline
cubic & \multicolumn{1}{r|}{97460 (\texttimes 1.16)}  & \multicolumn{1}{r|}{98}  & \multicolumn{1}{r|}{1611}  & \multicolumn{1}{r|}{14787617 (\texttimes 1.16)}  & \multicolumn{1}{r|}{96920 (\texttimes 1.16)}  & \multicolumn{1}{r|}{95}  & \multicolumn{1}{r|}{1608}  & \multicolumn{1}{r|}{14802517 (\texttimes 1.16)} \\ \hline
edn & \multicolumn{1}{r|}{3564 (\texttimes 1.29)}  & \multicolumn{1}{r|}{53}  & \multicolumn{1}{r|}{63}  & \multicolumn{1}{r|}{1157814 (\texttimes 1.22)}  & \multicolumn{1}{r|}{3944 (\texttimes 1.25)}  & \multicolumn{1}{r|}{53}  & \multicolumn{1}{r|}{64}  & \multicolumn{1}{r|}{1157585 (\texttimes 1.22)} \\ \hline
huffbench & \multicolumn{1}{r|}{4028 (\texttimes 1.39)}  & \multicolumn{1}{r|}{90}  & \multicolumn{1}{r|}{90}  & \multicolumn{1}{r|}{382404 (\texttimes 1.20)}  & \multicolumn{1}{r|}{3548 (\texttimes 1.33)}  & \multicolumn{1}{r|}{61}  & \multicolumn{1}{r|}{73}  & \multicolumn{1}{r|}{383163 (\texttimes 1.16)} \\ \hline
matmult-int & \multicolumn{1}{r|}{3376 (\texttimes 1.26)}  & \multicolumn{1}{r|}{29}  & \multicolumn{1}{r|}{70}  & \multicolumn{1}{r|}{1119255 (\texttimes 1.21)}  & \multicolumn{1}{r|}{2588 (\texttimes 1.23)}  & \multicolumn{1}{r|}{10}  & \multicolumn{1}{r|}{54}  & \multicolumn{1}{r|}{1200057 (\texttimes 1.21)} \\ \hline
minver & \multicolumn{1}{r|}{21472 (\texttimes 1.24)}  & \multicolumn{1}{r|}{56}  & \multicolumn{1}{r|}{489}  & \multicolumn{1}{r|}{151685 (\texttimes 1.18)}  & \multicolumn{1}{r|}{21404 (\texttimes 1.24)}  & \multicolumn{1}{r|}{59}  & \multicolumn{1}{r|}{474}  & \multicolumn{1}{r|}{178843 (\texttimes 1.17)} \\ \hline
nbody & \multicolumn{1}{r|}{20576 (\texttimes 1.20)}  & \multicolumn{1}{r|}{52}  & \multicolumn{1}{r|}{404}  & \multicolumn{1}{r|}{55767175 (\texttimes 1.18)}  & \multicolumn{1}{r|}{19944 (\texttimes 1.20)}  & \multicolumn{1}{r|}{42}  & \multicolumn{1}{r|}{387}  & \multicolumn{1}{r|}{55780098 (\texttimes 1.18)} \\ \hline
nettle-aes & \multicolumn{1}{r|}{5600 (\texttimes 1.14)}  & \multicolumn{1}{r|}{46}  & \multicolumn{1}{r|}{63}  & \multicolumn{1}{r|}{136279 (\texttimes 1.10)}  & \multicolumn{1}{r|}{5512 (\texttimes 1.14)}  & \multicolumn{1}{r|}{44}  & \multicolumn{1}{r|}{59}  & \multicolumn{1}{r|}{136447 (\texttimes 1.10)} \\ \hline
nettle-sha256 & \multicolumn{1}{r|}{7864 (\texttimes 1.07)}  & \multicolumn{1}{r|}{39}  & \multicolumn{1}{r|}{44}  & \multicolumn{1}{r|}{9573 (\texttimes 1.03)}  & \multicolumn{1}{r|}{7720 (\texttimes 1.08)}  & \multicolumn{1}{r|}{46}  & \multicolumn{1}{r|}{46}  & \multicolumn{1}{r|}{10239 (\texttimes 1.03)} \\ \hline
nsichneu & \multicolumn{1}{r|}{25204 (\texttimes 1.45)}  & \multicolumn{1}{r|}{654}  & \multicolumn{1}{r|}{565}  & \multicolumn{1}{r|}{3693 (\texttimes 1.44)}  & \multicolumn{1}{r|}{25152 (\texttimes 1.45)}  & \multicolumn{1}{r|}{648}  & \multicolumn{1}{r|}{546}  & \multicolumn{1}{r|}{3685 (\texttimes 1.44)} \\ \hline
qrduino & \multicolumn{1}{r|}{21840 (\texttimes 1.39)}  & \multicolumn{1}{r|}{554}  & \multicolumn{1}{r|}{455}  & \multicolumn{1}{r|}{1605197 (\texttimes 1.18)}  & \multicolumn{1}{r|}{18304 (\texttimes 1.37)}  & \multicolumn{1}{r|}{448}  & \multicolumn{1}{r|}{357}  & \multicolumn{1}{r|}{1610103 (\texttimes 1.16)} \\ \hline
slre & \multicolumn{1}{r|}{7604 (\texttimes 1.55)}  & \multicolumn{1}{r|}{242}  & \multicolumn{1}{r|}{211}  & \multicolumn{1}{r|}{--}  & \multicolumn{1}{r|}{6760 (\texttimes 1.52)}  & \multicolumn{1}{r|}{214}  & \multicolumn{1}{r|}{168}  & \multicolumn{1}{r|}{45480 (\texttimes 1.17)} \\ \hline
st & \multicolumn{1}{r|}{21964 (\texttimes 1.20)}  & \multicolumn{1}{r|}{58}  & \multicolumn{1}{r|}{420}  & \multicolumn{1}{r|}{6618952 (\texttimes 1.17)}  & \multicolumn{1}{r|}{21764 (\texttimes 1.19)}  & \multicolumn{1}{r|}{44}  & \multicolumn{1}{r|}{409}  & \multicolumn{1}{r|}{6626187 (\texttimes 1.17)} \\ \hline
statemate & \multicolumn{1}{r|}{8956 (\texttimes 1.29)}  & \multicolumn{1}{r|}{203}  & \multicolumn{1}{r|}{141}  & \multicolumn{1}{r|}{1573 (\texttimes 1.09)}  & \multicolumn{1}{r|}{9116 (\texttimes 1.28)}  & \multicolumn{1}{r|}{198}  & \multicolumn{1}{r|}{138}  & \multicolumn{1}{r|}{1672 (\texttimes 1.08)} \\ \hline
ud & \multicolumn{1}{r|}{3456 (\texttimes 1.25)}  & \multicolumn{1}{r|}{39}  & \multicolumn{1}{r|}{62}  & \multicolumn{1}{r|}{23674 (\texttimes 1.16)}  & \multicolumn{1}{r|}{3232 (\texttimes 1.27)}  & \multicolumn{1}{r|}{37}  & \multicolumn{1}{r|}{62}  & \multicolumn{1}{r|}{24125 (\texttimes 1.16)} \\ \hline
picojpeg & \multicolumn{1}{r|}{31116 (\texttimes 1.47)}  & \multicolumn{1}{r|}{881}  & \multicolumn{1}{r|}{665}  & \multicolumn{1}{r|}{2403701 (\texttimes 1.21)}  & \multicolumn{1}{r|}{22548 (\texttimes 1.49)}  & \multicolumn{1}{r|}{642}  & \multicolumn{1}{r|}{485}  & \multicolumn{1}{r|}{2424312 (\texttimes 1.16)} \\ \hline
sglib-combined & \multicolumn{1}{r|}{8332 (\texttimes 1.47)}  & \multicolumn{1}{r|}{197}  & \multicolumn{1}{r|}{219}  & \multicolumn{1}{r|}{409637 (\texttimes 1.18)}  & \multicolumn{1}{r|}{8736 (\texttimes 1.46)}  & \multicolumn{1}{r|}{206}  & \multicolumn{1}{r|}{214}  & \multicolumn{1}{r|}{466871 (\texttimes 1.31)} \\ \hline
wikisort & \multicolumn{1}{r|}{32340 (\texttimes 1.30)}  & \multicolumn{1}{r|}{354}  & \multicolumn{1}{r|}{647}  & \multicolumn{1}{r|}{28465027 (\texttimes 1.38)}  & \multicolumn{1}{r|}{31956 (\texttimes 1.29)}  & \multicolumn{1}{r|}{318}  & \multicolumn{1}{r|}{625}  & \multicolumn{1}{r|}{24817089 (\texttimes 1.20)} \\ \hline
\bf  geometric average & \multicolumn{1}{r|}{\bf  (\texttimes 1.30)}  & \multicolumn{1}{r|}{}  & \multicolumn{1}{r|}{}  & \multicolumn{1}{r|}{\bf  (\texttimes 1.19)}  & \multicolumn{1}{r|}{\bf  (\texttimes 1.29)}  & \multicolumn{1}{r|}{}  & \multicolumn{1}{r|}{}  & \multicolumn{1}{r|}{\bf  (\texttimes 1.18)} \\ \hline
\end{tabular}
\end{table*}

\begin{figure*}
  \centerline{%
    \includegraphics[width=.495\linewidth]{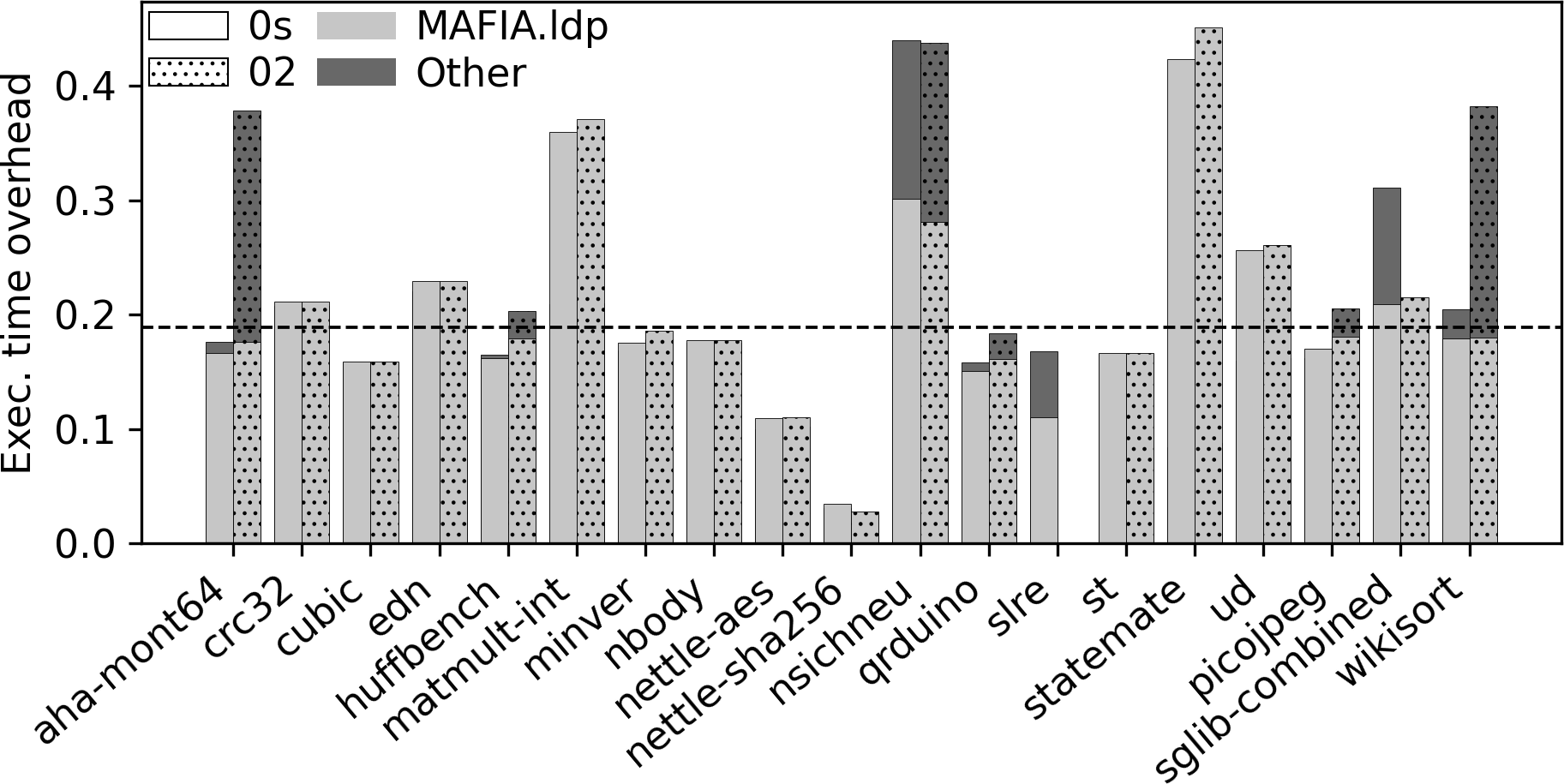}
  \hspace*{\fill{}}
    \includegraphics[width=.495\linewidth]{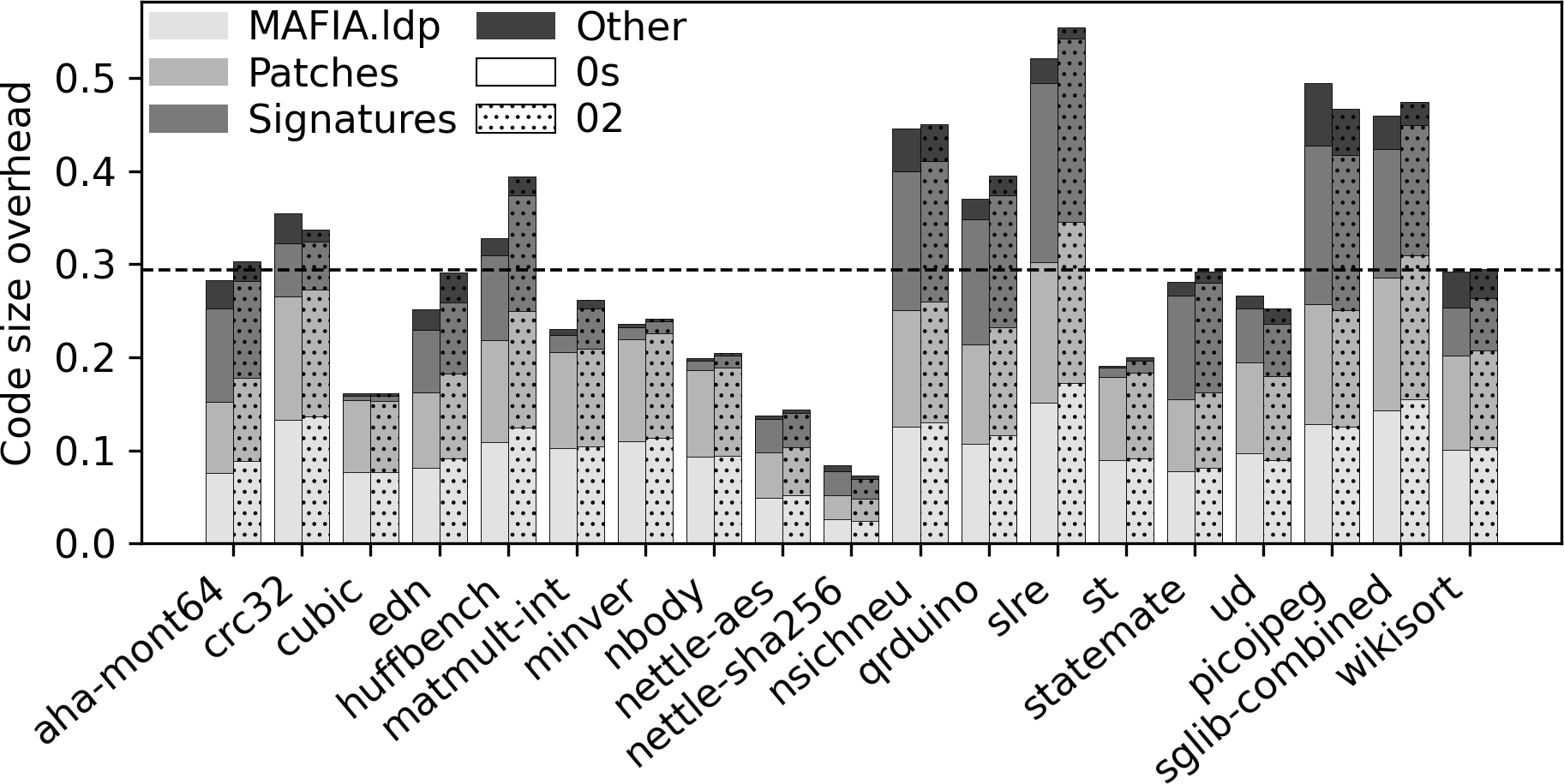}%
    }
  \caption{Execution time (left) and code size (right) overheads for the Embench-IoT benchmarks,
    with the {\tt{}-Os} 
    and {\tt{}-O2} 
    compiler optimization levels
    for \name{} with the CRC signature function.
  }
  \label{fig:overheads}
\end{figure*}
Table~\ref{table:result} and Figure~\ref{fig:overheads} summarize the results of our experimental evaluation.
The results show that \name{} can handle different kind of software as selected by Embench-IOT.
Execution time overheads range between 2.5\% and 44.0\% with a geometric average of 18.4\%.
The code size overheads range between 7.3\% and 55.4\% with a geometric average of 29.4\%.
For the majority of the benches, the difference between the optimization levels \code{-Os} and \code{-O2} is negligible,
except for \code{aha-mont64}, \code{sglib-combined}, and \code{wikisort}.
The difference is explained by the execution of instruction sequences where a \code{\name{}.ldp} is immediately followed by a branch.
In this case, the pipeline controller stalls the processor so that the memory stage can fetch the patch before the branch is taken.
For those 3 benches, this pattern is present in a loop nest increasing considerably the number of stalled cycles between optimization level \code{-O2} and \code{-Os}.

The evaluation results show that the update function (patch values and \code{\name{}.ldp}) is responsible for the largest part of the code size overhead.
The number of reference signatures is dependent of the number of branches in the functions annotated with \code{\name{}\_secured}.
In our evaluations, reference signatures are only inserted for the benches core functions.
But, all the functions need \code{\name{}.ldp} independently whether they come from the benchmark core or from libraries.

The forwarding dependency elimination pass and the patch placement pass (Section~\ref{sec:software_support}) also contribute to the code size overhead (``Other'' in Fig.~\ref{fig:overheads}).
Those passes insert \code{nop} instructions to break some forwarding dependencies and disable some tail call optimization respectively.
Also, \name{} LLVM passes insert new instructions in the code (e.g. \code{\name{}.ldp} instructions).
This requires to adapt direct branch offsets.
The new offsets might not fit in the original branch binary encoding anymore.
In such a case, an additional direct branch is inserted which increases the code size.

Table~\ref{table:dispatchers_overhead} reports the overhead induced by the dispatchers.
In the worst case, the dispatchers contribute to 15.3\% of the total overhead for \code{wikisort -Os}. The absolute overhead induced is only 4\% at worst.
For all the other benches, the dispatchers contribute to less than 10\% of the total overhead.
Note that this overhead could even be reduced by replacing dispatcher targeting equivalence classes with a single element to a direct call to the target function.
This demonstrates that using dispatchers is a practical solution to avoid GPSA signature confusion for a small additional overhead.

\begin{table}[]
\centering
\caption{Contribution of dispatchers to the total code size overhead: in bytes, 
number of added patches and added signatures}
\label{table:dispatchers_overhead}
\begin{tabular}{|l|r|r|r|}
\hline
Bench & Code Size (Bytes, \%) & Patches & Signatures \\
\hline
picojpeg (\code{Os})        & 84 (1.1\%)    & 4  & 2   \\
\hline
picojpeg (\code{O2})        & 168  (1.7\%)   & 4  & 4  \\
\hline
sglib-combined (\code{Os})  & 256  (9.3\%)   & 8  & 6  \\
\hline
sglib-combined (\code{O2})  & 172  (6.4\%)   & 6  & 4  \\
\hline
wikisort (\code{Os})        & 1108  (15.3\%) & 34 & 26 \\
\hline
wikisort (\code{O2})        & 604  (8.2\%)   & 22 & 14  \\
\hline
\end{tabular}
\end{table}

We observe that the compiler optimization level has a moderate impact on the code size and execution time overheads,
but that there is a large variation of overheads,
which are due to the different code structures used in the benches.
Furthermore, smaller basic block sizes are more impacted by the instrumentation with \name{} instructions.
The maximum code size overhead is 400\% ($\times4$) for a single basic block composed of a single branch instruction,
as
\name{}'s code instrumentation requires the addition of 3 memory words: 
a \code{\name{}.ldp} instruction, 
the corresponding patch value, and the reference signature.
Our results show that the average code size overhead is far less because basic blocks have greater sizes. 
Some optimizations (such as tail duplication or loop unrolling) could increase the size of basic blocks or reduce the number of branches to reduce the execution time overheads.

{%
It is possible to get an approximation of the CBC-MAC/Prince overheads from the CRC32 results.
Both CBC-MAC/Prince and CRC32 compute a signature in a single cycle and verify 32-bit reference signature.
However, CBC-MAC/Prince works with 64-bit blocks and therefore requires 64-bit patch values.
To handle 64-bit patch values \name{} uses two update instructions, one for the 32 most significant bits and the other for the 32 least significant bits.
Such implementation leads to double the software overheads induced by the update function (patch values and \texttt{\name{}.ldp}).
Therefore, with the CBC-MAC/Prince implementation, \name{} induces an average execution time overhead close to 39\% and an average code size overhead close to 50\%,
but it also ensures code authenticity.
Replacing Prince with a 
32-bit block cipher, such as Simon~\cite{MaeneSingleCycleImplementations2016}, could close the performance gap between the CBC-MAC and the CRC32 implementations of \name{}.
}

\section{Related Work}\label{sec:related work}
Counter-measures ensuring code and control-flow integrity are often implemented as hardware components external to the processor microarchitecture~\cite{aroraHardwareAssistedRunTimeMonitoring2006,dangerCCFICacheTransparentFlexible2018}.
Such counter-measures are easier to integrate in a processor design
but are intrinsically blind to faults targeting the microarchitecture.  
{For these reasons, they are not further discussed in this section.}

{%
  Table~\ref{tab:comparative} provides a comparison between \name{} and the related counter-measures.
  It reports the claimed security properties ensured by the counter-measures, the estimated hardware area,  code size execution time overhead.
  Regarding the hardware area overhead, note that Table~\ref{tab:comparative} is only indicative of a trend because each work is based on different processor architectures and different technologies.%
  }

{%
Werner et al.\ use GPSA in the context of fault injection~\cite{wernerProtectingControlFlow2015}.
The signature is derived from the binary encoding of program instructions using a CRC32 signature function.
In \name{}, the CRC32 implementation presents slightly larger hardware overhead due to the additional CSI module.
Actually, the CACFI module, which handles GPSA in \name{},  is equivalent to the GPSA monitor in \cite{wernerProtectingControlFlow2015}.
The main difference is the origin of the signature input which in \name{} is the pipeline state instead of the binary encoding of program instructions.
On the software side, \name{} induces half less code size overhead and execution time overhead.
The reason is that \name{} handles the loading of patch values in a single dedicated \texttt{\name{}.ldp} instruction while \cite{wernerProtectingControlFlow2015} requires several standard loads and stores to place the patch values in a memory mapped register.
Regarding the security properties, both \name{} and \cite{wernerProtectingControlFlow2015} ensure code and control-flow integrity.
Additionally, \name{} supports \revised{control-signal integrity}, meaning that it can detect fault targeting the control signals after the fetch stage.
}

\cite{clercqSOFIASoftwareControl2016, wernerSpongeBasedControlFlowProtection2018, savryConfidaentControlFLow2020} are code and control-flow integrity counter-measures based on authenticated decryption.
\name{} ensures code authenticity, but is not designed to ensure code confidentiality as is.
\cite{wernerSpongeBasedControlFlowProtection2018} is the closest to
\name{} implemented with a CBC-MAC signature function.
Both designs have similar hardware overheads.
They are both based on the Prince cryptographic primitive, which is responsible for the most important usage of extra silicon area.
However, the software overheads of \cite{wernerSpongeBasedControlFlowProtection2018} are approximately 30\% smaller for two reasons:
(i)~faults are detected in case of bad instruction decoding, which alleviates the need for reference signatures;
(ii)~control-flow instructions simultaneously load patch values, whereas \name{} requires a dedicated instruction.

Regarding indirect control flow, \cite{clercqSOFIASoftwareControl2016} also relies on indirect branch elimination.
By design, the basic blocks cannot have more than 4 or 5 instructions, and cannot have more than 2 predecessors.
Those limitations considerably increase the complexity of the dispatchers, which leads to larger software overheads as compared \name{}.
In \cite{wernerSpongeBasedControlFlowProtection2018} all the functions reachable from the same indirect branch share the same state, which is similar to signature confusion.
However, the state is also used for code encryption.
To avoid possible cryptographic vulnerabilities, an additional patch value is inserted at the beginning of the target functions, which restricts possible signature confusions to the function entry points.
In \cite{savryConfidaentControlFLow2020}, indirect branch patch values are stored in a word placed before the indirect branch target.
This approach prevents any signature confusion and allows more flexibility to support software constructs such as C++ \code{vtable}s.
However, this mechanism requires confidential patch values to prevent the forging of new indirect branches.
\name{} relies on indirect branch elimination to prevent signature confusion.
We evaluated that the overheads due to the use of dispatchers are low,
and that they could be further reduced with additional code optimizations in the toolchain.
Finally,
all the related works in the context of control-flow integrity against fault injection attacks implicitly assume that indirect branch targets can be identified by external means.
This is a major bottleneck for the use of any counter-measure in practice.
It implies either
that the burden is moved to another tool or to an application designer,
or in the worst case the use of a single equivalence class containing all the targets of indirect branches.
Our work is the only one to provide a fully automated solution for indirect branch target identification.

\begin{table*}[]
\centering
\caption{Security and overhead comparison of code and control flow integrity protection targeting fault injection attacks}
\label{tab:comparative}
\begin{tabular}{|l|l|l|l|r|r|r|}
\hline
 &
  Security &
  \multicolumn{1}{c|}{Target} &
  \multicolumn{1}{c|}{Technology} &
  \multicolumn{1}{c|}{\begin{tabular}[c]{@{}c@{}}Area\\ overhead\end{tabular}} &
  \multicolumn{1}{c|}{\begin{tabular}[c]{@{}c@{}}Exec. time\\ overhead\end{tabular}} &
  \multicolumn{1}{c|}{\begin{tabular}[c]{@{}c@{}}Code size\\ overhead\end{tabular}} \\ \hline
  Arora et al. \cite{aroraHardwareAssistedRunTimeMonitoring2006}  & CI/CFI          & ARM9TDMI ARM920T   & FPGA Virtex 2   & 13.7\% & 100\%       & NA             \\ \hline
  Werner et al. \cite{wernerProtectingControlFlow2015}            & CI/CFI          & ARMv7-M Cortex-M3 & ASIC UMC 130nm  & 4\%    & 32\%        & 57\%           \\ \hline
  Danger et al. \cite{dangerCCFICacheTransparentFlexible2018}          & CI/CFI          & RISC-V PicoRV32  & FPGA Artix 7    & 20\%   & 2\% to 63\% & 118\% to 160\% \\ \hline
  Clercq et al. \cite{clercqSOFIASoftwareControl2016}             & CC/CA/CFI       & SPARC LEON3     & FPGA Virtex 6   & 28.2\% & 13.7\%      & 140\%          \\ \hline
  Werner et al. \cite{wernerSpongeBasedControlFlowProtection2018} & CC/CA/CFI       & RISC-V CV32E40P  & ASIC UMC 65nm   & 28.8\% & 9.1\%       & 19.8\%         \\ \hline
Savry et al. \cite{savryConfidaentControlFLow2020}                & CC/CA/CFI/DC/DA & RISC-V CV32E40P  & --               & --      & 167\%       & 24\%           \\ \hline
MAFIA CRC                                                         & CI/CFI/\revised{CSI}       & RISC-V CV32E40P  & ASIC FDSOI 22nm & 6.5\%  & 18.4\%      & 29.4\%         \\ \hline
MAFIA CBC-MAC                                                     & CA/CFI/\revised{CSI}       & RISC-V CV32E40P  & ASIC FDSOI 22nm & 23.8\% & 39\%        & 50\%           \\ \hline
\end{tabular}

\smallskip
CI:~Code Integrity, CA:~Code Authenticity, CC:~Code Confidentiality, CFI:~Control Flow Integrity, DC:~Data Confidentiality, DA:~Data Authenticity, \revised{\revised{CSI}:~Control-Signal Integrity}, NA:~Non-Applicable, --:~Not provided
\end{table*}

{%
  While \name{} has comparable overheads to the related counter-measures, it is still possible to reduce those overheads.
  First, the area overhead is estimated considering the overhead on the processor core only.
  However, the CV32E40P is a small processor core and we expect the contribution of the core to the area of a full system to be small, even in the case of IoT devices.
Hence, \name{}'s relative area overhead measured in a complete system, including memory and peripherals, would be much smaller.
  It is also possible to reduce \name{}'s area overhead and code size overhead, at the expense of security, by selecting a more lightweight signature function such as CRC8.
  Finally, our work reports \name{}'s code size and execution time overheads when applied globally to the application.
  By extending \name{} with a secure on/off mechanism, it would be possible to enable \name{} protection for sensitive code only.
  The local overheads on the sensitive code would be the same as the ones reported in Section~\ref{sec:experimental_evaluation}, but the global overheads on the application would be much smaller. %
}

\name{} is, to the best of our knowledge, the only counter-measure to ensure \revised{control-signal integrity} against fault injection attacks.
Kim and Somani propose an on-line integrity monitoring of the microprocessor control logic for safety-critical systems sensible to soft errors~\cite{kimOnlineIntegrityMonitoring2001}.
They use a non-secure function (XOR) to derive a signature from the static control signals in every pipeline stages, and dynamic control signals are protected by duplication.
A caching mechanism is used to store reference signatures from the first execution (cache miss), and verifies the runtime signatures in the subsequent executions (cache hit).
{However, such a caching mechanism does not protect against attacks targeting executions leading to a signature cache miss (e.g.\ first program execution) because the reference signature is not available}.
Moreover, this solution does not detect attacks targeting the program memory, hence does not ensure code integrity.
Another independent mechanism ensures control-flow integrity.
In \name{}, a unique signature is derived from static and dynamic control signals of the decode stage.
This signature ensures simultaneously execution, code and control-flow integrity.

\section{Conclusion}\label{sec:conclusion}
This paper presents \name{}, a counter-measure extending the state-of-the-art against fault injection attacks by combining \revised{control-signal integrity} with code integrity, code authenticity and control-flow integrity.
\name{} articulates two security mechanisms to protect the control logic of the processor against faults targeting the processor microarchitecture.
\name{} also protects against faults injected outside of the processor that have an impact on the processor control logic.
A first module implements generalized path signature analysis (GPSA).
The signature is computed from the pipeline state, a set of data-independent control signals that {deterministically result from the decoding of the binary instruction}.
This module ensures simultaneously control-flow integrity, code authenticity, and control signal integrity from the fetch stage to the end of the decode stage.
A second module, implementing a redundancy-based mechanism, ensures the integrity of the same control signals in the subsequent pipeline stages, which completes the full protection coverage of the processor microarchitecture.

{
The flexibility of the design allows for trade-offs between security and overheads.
The paper presents two implementations of \name{} based on the CV32E40P RISC-V processor, with different signature functions:
one with CBC-MAC and Prince, and another one with a CRC32 error detector code.
CBC-MAC/Prince makes use of the full capabilities of \name{}.
It induces a hardware area overhead of 23.8\,\%, and average code size and an execution time overheads of 50\,\% and 39\,\% respectively.
CRC32 detects a minimum number of 8 bit-flips and ensures code integrity only instead of code authenticity; it induces a hardware area overhead of 6.5\,\%, and average code size and an execution time overheads of 
29.4\,\% and 18.4\,\% respectively.
On the software side, the compiler extension offers a complete automatic processing of the program source code to generate the \name{} executable program.
Moreover, thanks to the support of indirect branches and interrupts, \name{} is fully compliant with software stacks used in embedded system.
}

\section*{Acknowledgements}
We thank Mikael Le Coadou and Juan Suzano Da Fonseca for their contributions to the hardware evaluation, and Simon Tollec for his contribution to the formal verification.

\bibliographystyle{IEEEtran}
\bibliography{biblio,damien}

\begin{IEEEbiography}[{\includegraphics
[width=1in,height=1.25in,clip,
keepaspectratio]{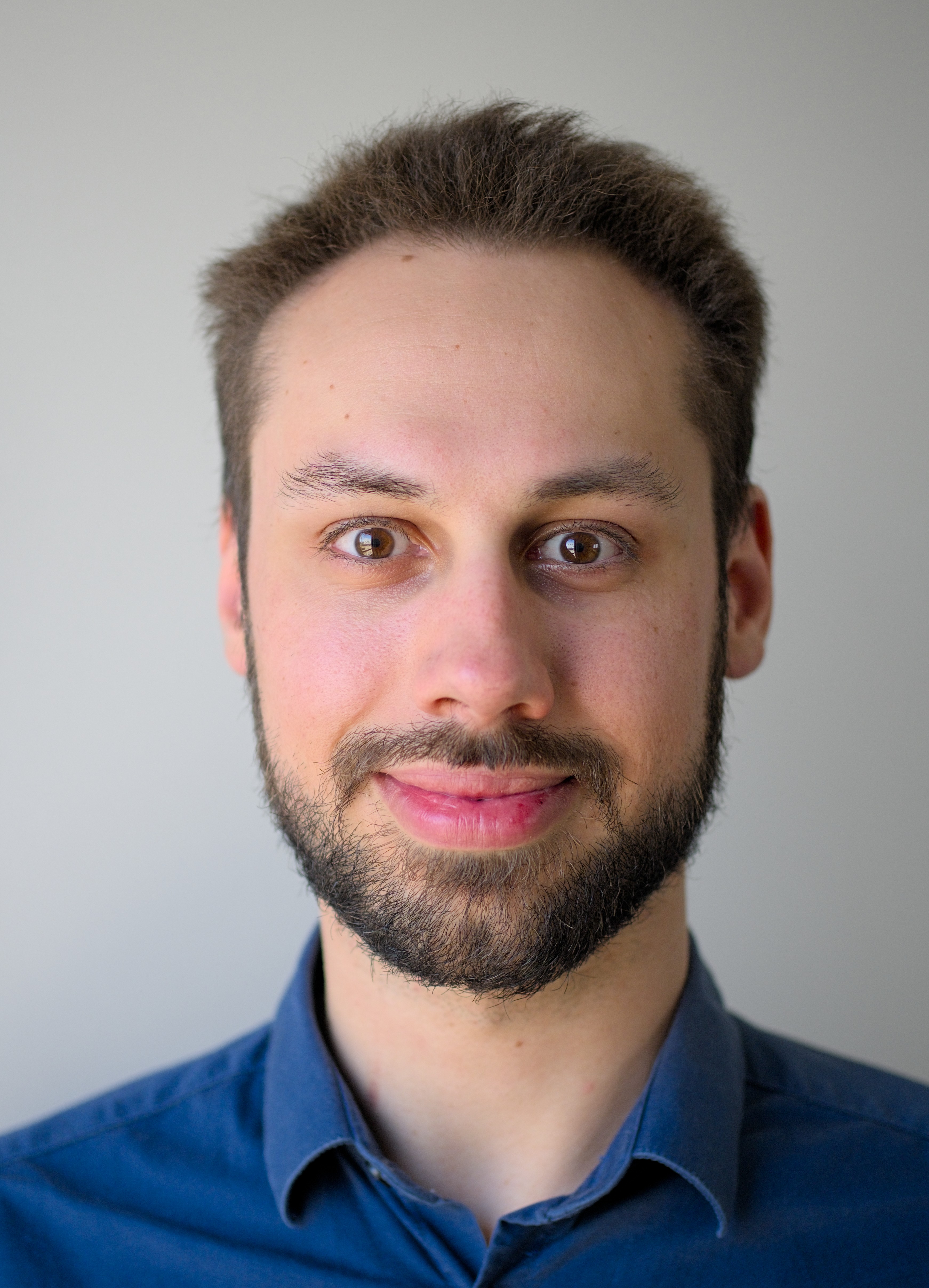}}]
{Thomas Chamelot}
received the M.S.\ degree in Cybersecurity from the University Toulouse III, Toulouse, France in 2019, and  the Ph.D. degree from Sorbonne Univeristé, Paris, France, in 2022.
His current research interests include hardware security and computer architecture.

\end{IEEEbiography}

\begin{IEEEbiography}[{\includegraphics
[width=1in,height=1.25in,clip,
keepaspectratio]{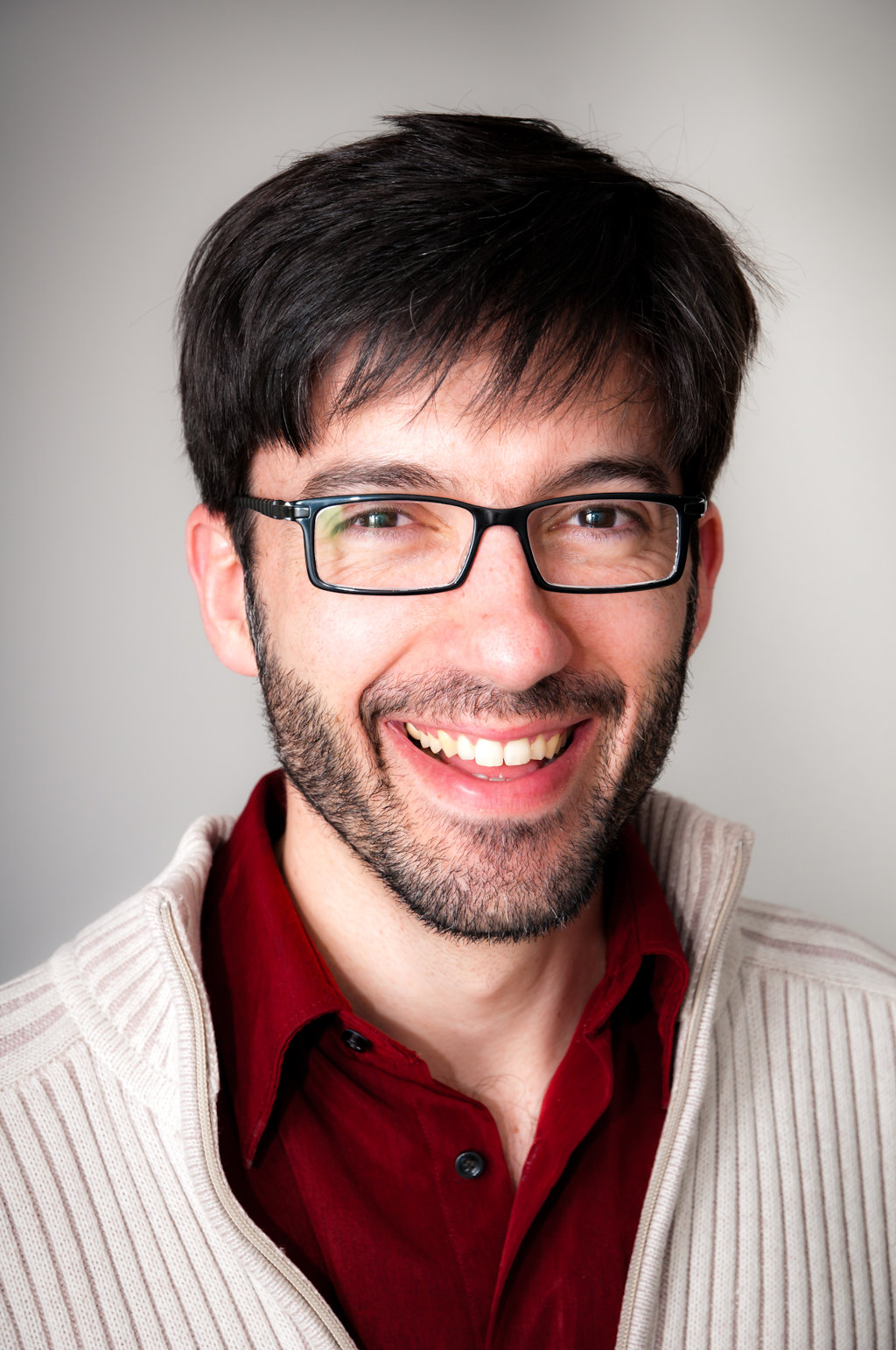}}]
{Damien Couroussé}
is with CEA-List since 2011, as a Research Engineer and Senior Expert.
He received the Ph.D.\ from the Institut National Polytechnique de Grenoble in 2008.
His research interests include embedded software and its interaction with hardware, compilation and runtime code generation for performance and security, with a recent focus on hardware security.
\end{IEEEbiography}

\begin{IEEEbiography}[{\includegraphics
[width=1in,height=1.25in,clip,
keepaspectratio]{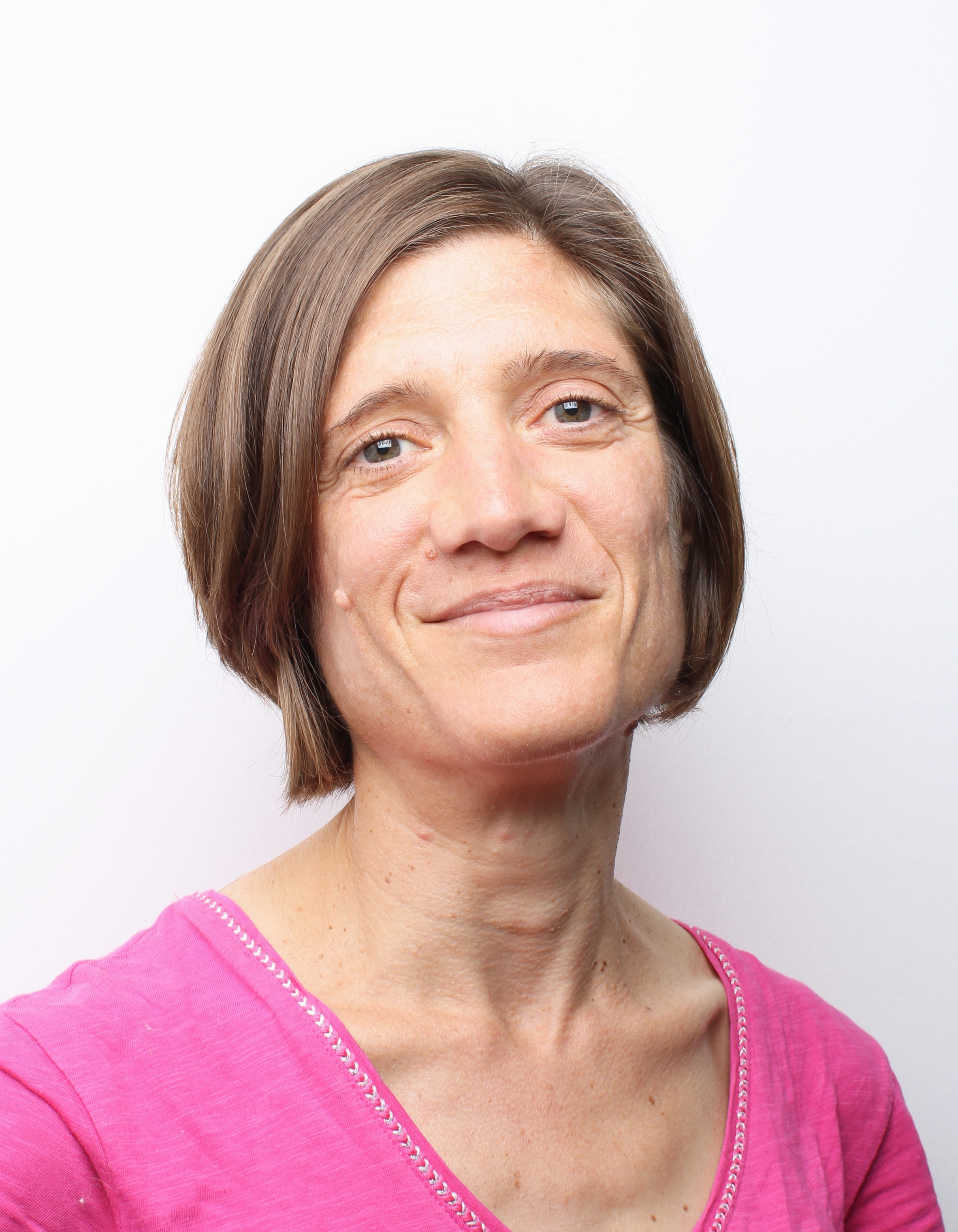}}]
{Karine Heydemann}
is an Associate Professor
at Sorbonne University / LIP6 since 2006 and a Senior Expert Architect at Thales DIS since September 2022.
She received the Ph.D.\ degree in Computer Science
from the University of Rennes 1 in 2004.
Her areas of expertise encompass hardware
micro-architecture, compilation, code optimization,
and physical attacks, including modelling of
hardware fault injection effects, automated code
hardening and robustness analysis.

\end{IEEEbiography}

\newpage
\vfill
\listoftodos

\end{document}